\documentclass[12pt]{article}

\usepackage{epsf}
\usepackage{rotate}

\newcommand{\lsim}{
\mathrel{\hbox{\rlap{\hbox{\lower4pt\hbox{$\sim$}}}\hbox{$<$}}}}
\newcommand{\gsim}{
\mathrel{\hbox{\rlap{\hbox{\lower4pt\hbox{$\sim$}}}\hbox{$>$}}}}

\begin{document}

\begin{titlepage}

\begin{flushright}
CERN-TH/2003-011\\
hep-ph/0301256
\end{flushright}

\vspace{2cm}
\begin{center}
\boldmath
\large\bf A Closer Look at $B_{d,s}\to D f_r$ Decays and

\vspace{0.3truecm}

Novel Avenues to Determine $\gamma$
\unboldmath
\end{center}

\vspace{1.2cm}
\begin{center}
Robert Fleischer\\[0.1cm]
{\sl Theory Division, CERN, CH-1211 Geneva 23, Switzerland}
\end{center}

\vspace{1.7cm}
\begin{abstract}
\vspace{0.2cm}\noindent
Decays of neutral $B$ mesons originating from $b\to D r$ $(r\in\{s,d\})$ 
processes, i.e.\ $B_d\to DK_{\rm S(L)}$, $B_s\to D\eta^{(')}, D\phi, ...$\ 
$(r=s)$ and $B_s\to DK_{\rm S(L)}$, $B_d\to D\pi^0, D\rho^0, ...$\ $(r=d)$
modes, offer valuable insights into CP violation. Here the neutral $D$ mesons 
may be observed through $D\to f_\pm$ or $D\to f_{\rm NE}$, where $f_\pm$ and 
$f_{\rm NE}$ are CP eigenstates and CP non-eigenstates, respectively. Recently,
we pointed out that ``untagged'' and mixing-induced CP-violating $B_{d,s}\to 
D_\pm f_s$ observables provide a very efficient, theoretically clean 
determination of the angle $\gamma$ of the unitarity triangle. Here we perform
a more detailed analysis of the $B_{d,s}\to D[\to f_\pm]f_s$ channels, and 
focus on $B_{d,s}\to D[\to f_{\rm NE}] f_s$, where interference effects 
between Cabibbo-favoured and doubly Cabibbo-suppressed $D$-meson decays lead 
to complications. We show that $\gamma$ can nevertheless be determined 
in an elegant and essentially unambiguous manner from the 
$B_{d,s}\to D[\to f_{\rm NE}]f_s$ observables with the help of the
corresponding ``untagged'' $B_{d,s}\to D[\to f_\pm]f_s$ measurements. 
Moreover, we may also extract hadronic $B$- and $D$-decay parameters. 
Several interesting features of decays of the kind $B_{d,s}\to D f_d$ are 
also pointed out.
\end{abstract}

\vfill
\noindent
CERN-TH/2003-011\\
January 2003

\end{titlepage}

\thispagestyle{empty}
\vbox{}
\newpage
 
\setcounter{page}{1}

\section{Introduction}\label{intro}
\setcounter{equation}{0}
The exploration of CP violation, which was discovered in 1964 through
the observation of $K_{\rm L}\to\pi^+\pi^-$ decays \cite{CP-obs}, is 
one of the hot topics in particle physics. In this decade, studies of 
$B$-meson decays will provide various stringent tests of the 
Kobayashi--Maskawa mechanism of CP violation \cite{KM}, allowing us to 
accommodate this phenomenon in the Standard Model of electroweak 
interactions. At present, the experimental stage is governed by the 
asymmetric $e^+e^-$ $B$ factories operating at the $\Upsilon(4S)$ 
resonance, with their detectors BaBar (SLAC) and Belle (KEK). Interesting
$B$-physics results are also soon expected from run II of the Tevatron
\cite{TEV-II}. Several strategies can only be fully exploited in the era 
of the LHC \cite{LHC-Report}, in particular by LHCb (CERN) and BTeV 
(Fermilab).

\begin{figure}[h]
\vspace*{0.3truecm}
\begin{center}
\leavevmode
\epsfysize=5.0truecm 
\epsffile{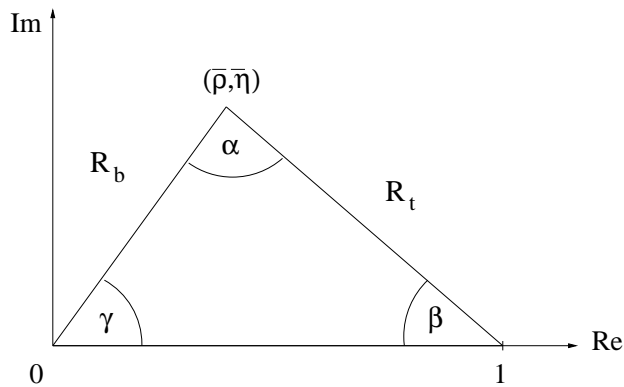} 
\end{center}
\caption{Unitarity triangle of the CKM matrix.}\label{fig:UT}
\end{figure}

The central target for the exploration of CP violation is the unitarity 
triangle of the Cabibbo--Kobayashi--Maskawa (CKM) matrix illustrated in 
Fig.~\ref{fig:UT}, which can be determined through measurements of its 
sides on the one hand, and direct measurements of its angles through 
CP-violating effects in $B$ decays on the other hand (for a detailed review, 
see \cite{RF-PHYS-REP}). If we take into account next-to-leading order 
terms in the Wolfenstein expansion of the CKM matrix \cite{wolf}, 
following the prescription given in \cite{BLO}, we obtain the following
coordinates for the apex of the unitarity triangle:
\begin{equation}
\overline{\rho}\equiv(1-\lambda^2/2)\rho, \quad
\overline{\eta}\equiv(1-\lambda^2/2)\eta,
\end{equation}
where $\lambda\equiv|V_{us}|=0.22$. 
The goal is to overconstrain the unitarity triangle as much as possible. 
In this respect, a first important milestone has already been achieved by 
the BaBar and Belle collaborations, who have announced the observation of 
CP violation in the $B$-meson system in the summer of 2001 \cite{CP-B-obs}.
This discovery was made possible by $B_d\to J/\psi K_{\rm S}$ and similar 
channels. As is well known, these modes allow us to determine $\sin\phi_d$, 
where $\phi_d$ denotes the CP-violating weak $B^0_d$--$\overline{B^0_d}$ 
mixing phase, which is given by $2\beta$ in the Standard Model. Using the 
present world average $\sin\phi_d=0.734\pm0.054$ \cite{nir}, we obtain the 
following twofold solution for $\phi_d$ itself:
\begin{equation}
\phi_d=\left(47^{+5}_{-4}\right)^\circ \, \lor \, 
\left(133^{+4}_{-5}\right)^\circ,
\end{equation}
where the former would be in perfect agreement with the ``indirect'' range 
implied by the CKM fits, $40^\circ\lsim\phi_d\lsim60^\circ$ \cite{CKM-fits}, 
while the latter would correspond to new physics. The two solutions can 
be distinguished through a measurement of the sign of $\cos\phi_d$.
Several strategies to accomplish this important task were proposed 
\cite{ambig}. For example, in the $B\to J/\psi K$ system, 
$\mbox{sgn}(\cos\phi_d)$ can be extracted with the help of the 
time-dependent angular distribution of the $B_d\to J/\psi[\to\ell^+\ell^-] 
K^\ast[\to\pi^0K_{\rm S}]$ decay products if the sign of a hadronic parameter 
$\cos\delta_f$, involving a strong phase $\delta_f$, is fixed through 
factorization \cite{DDF2,DFN}. This analysis is already in progress at the 
$B$ factories \cite{itoh}. A remarkable link between the two solutions 
for $\phi_d$ and ranges for $\gamma$ is also suggested by recent data on 
the CP-violating $B_d\to\pi^+\pi^-$ observables \cite{FlMa2}.
 
Interesting tools for the exploration of CP violation are provided by
$B_d\to DK_{\rm S(L)}$, $B_s\to D\eta^{(')}, D\phi, ...$\ and 
$B_s\to DK_{\rm S(L)}$, $B_d\to D\pi^0, D\rho^0, ...$\ modes, which 
originate from $b\to Ds$ and $b\to D d$ transitions, respectively, and 
receive only contributions from tree-diagram-like topologies 
\cite{GroLo}--\cite{AtSo}. The neutral $D$ mesons may be observed 
through decays into CP eigenstates $f_\pm$, or through decays into CP 
non-eigenstates $f_{\rm NE}$. In a recent paper \cite{RF-BD-CP}, we pointed 
out that $B_d\to D_\pm K_{\rm S (L)}$ and 
$B_s\to D_\pm \eta^{(')}, D_\pm \phi, ...$\ 
decays, where $D_+$ and $D_-$ denote the CP-even and CP-odd eigenstates 
of the neutral $D$-meson system, respectively, offer very efficient, 
theoretically clean determinations of the angle $\gamma$ of the unitarity 
triangle. In this strategy, only ``untagged'' and 
mixing-induced CP-violating observables are employed, which satisfy a very 
simple relation, allowing us to determine $\tan\gamma$. Using a plausible 
dynamical assumption, $\gamma$ can be fixed in an essentially unambiguous 
manner. The corresponding formalism can also be applied to 
$B_d\to D_\pm\pi^0,D_\pm\rho^0, ...$\ and $B_s\to D_\pm K_{\rm S (L)}$ 
decays. Although these modes appear less attractive for the extraction of 
$\gamma$, they provide interesting determinations of $\sin\phi_q$. In 
comparison with the conventional $B_d\to J/\psi K_{\rm S(L)}$ and 
$B_s\to J/\psi\phi$ methods, these extractions do not suffer from any 
penguin uncertainties, and are theoretically cleaner by one order of 
magnitude.

In the present paper, we give a more detailed discussion of the $D_\pm$ 
case by having a closer look at $D\to f_\pm$ processes, and focus on the 
CP non-eigenstate $D\to f_{\rm NE}$ case. Here we have to deal with 
interference effects between Cabibbo-favoured 
and doubly Cabibbo-suppressed $D$-meson decays, which lead to certain 
complications \cite{KaLo,ADS}. We point out that $\gamma$ can nevertheless 
be determined in a powerful and essentially unambiguous way from the 
time-dependent $B_d\to D[\to f_{\rm NE}]K_{\rm S(L)}$, 
$B_s\to D[\to f_{\rm NE}]\eta^{(')}, D[\to f_{\rm NE}]\phi, ...$\ rates with 
the help of simple ``untagged'' measurements of the corresponding 
$B_d\to D_\pm K_{\rm S (L)}$, $B_s\to D_\pm \eta^{(')}, D_\pm \phi, ...$ 
processes. Moreover, we may also extract interesting hadronic $B$- and 
$D$-decay parameters. This strategy can be implemented with the help of
transparent formulae in an {\it analytical} manner. Decays of the kind 
$B_s\to D K_{\rm S (L)}$, $B_d\to D\pi^0,D\rho^0, ...$\ also exhibit 
interesting features, but they appear not to be as attractive for the 
extraction of $\gamma$ as their $b\to D s$ counterparts, since they have 
a different CKM structure. 

For the testing of the Kobayashi--Maskawa mechanism of CP violation 
\cite{KM}, it is essential to measure $\gamma$ in a variety of ways. 
In particular, it will be very interesting to compare the theoretically 
clean values for $\gamma$ obtained with the help of the new strategies 
proposed here with the results that will be provided by other approaches. 
In this context, $B\to\pi K$ and $B_s\to K^+K^-$, $B_d\to\pi^+\pi^-$ modes 
are important examples, allowing the extraction of $\gamma$ through $SU(3)$ 
flavour-symmetry arguments (for overviews, see \cite{BpiK-overviews}).
Several theoretically clean strategies were also proposed, employing various 
different pure tree-diagram-like $B$ decays (see, for instance, \cite{ADS} 
and \cite{GW}--\cite{gro-BDK-02}). In general, these approaches are either 
conceptually simple but experimentally very challenging, or do not allow 
transparent extractions of $\gamma$ through simple analytical expressions, 
thereby also complicating things considerably and yielding multiple discrete 
ambiguities. Alternative approaches to extract weak phases from neutral
$b\to D s$ modes can be found in \cite{GroLo}--\cite{AtSo}.

The outline of this paper is as follows: in Section~\ref{sec:Evol-Ampl}, we 
introduce a convenient notation and discuss the time evolution and amplitude 
structure of neutral $B$-meson decays caused by $b\to Ds$ and $b\to Dd$ 
transitions. In Section~\ref{sec:CP}, we then turn to the case where the 
neutral $D$ mesons are observed through decays into CP eigenstates $f_\pm$,
whereas we shall focus on the interesting opposite case, where the $D$ mesons 
are observed through decays into CP non-eigenstates $f_{\rm NE}$, in 
Section~\ref{sec:CPNE}. The main approach of this paper is presented in 
Section~\ref{sec:strat}, and is also illustrated there with the help of 
a few instructive examples. Finally, we conclude in Section~\ref{sec:concl}.

\begin{figure}
\begin{center}
\leavevmode
\epsfysize=5.0truecm 
\epsffile{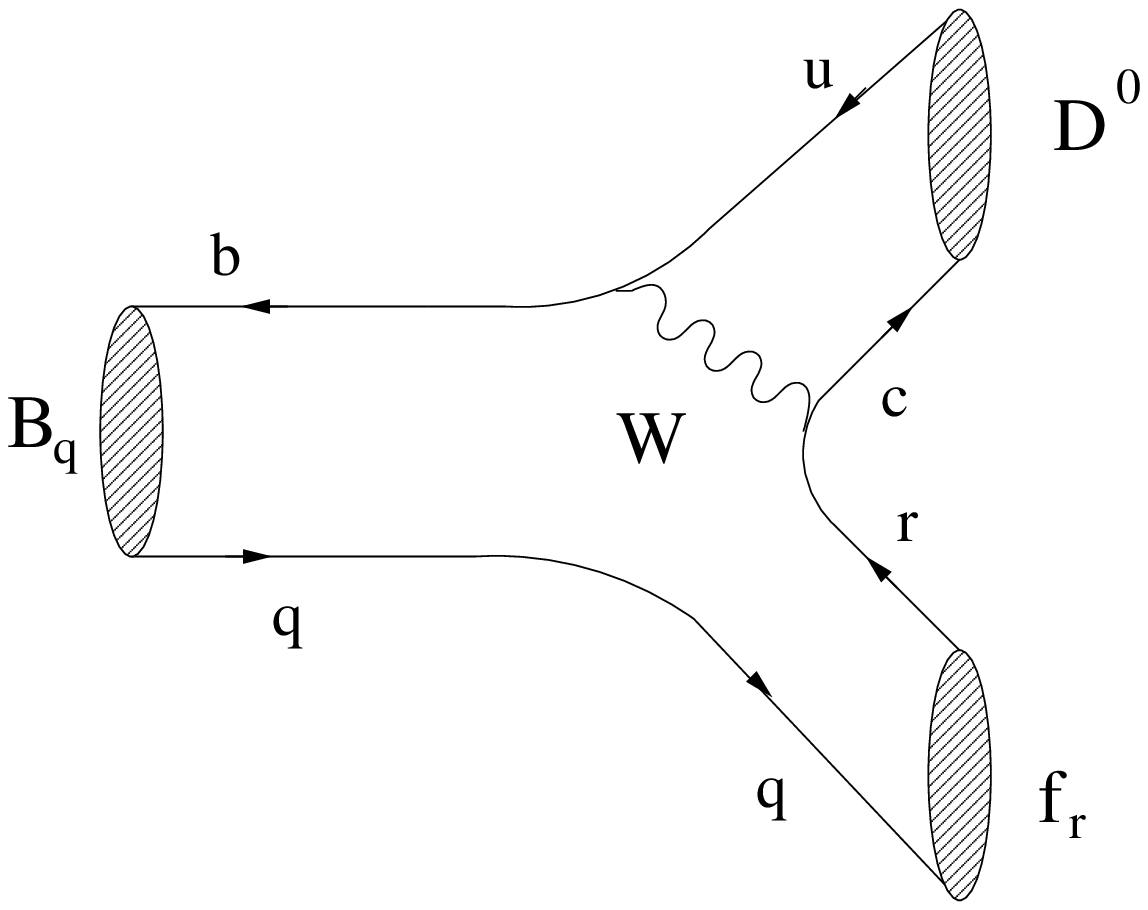} \hspace*{1truecm}
\epsfysize=5.0truecm 
\epsffile{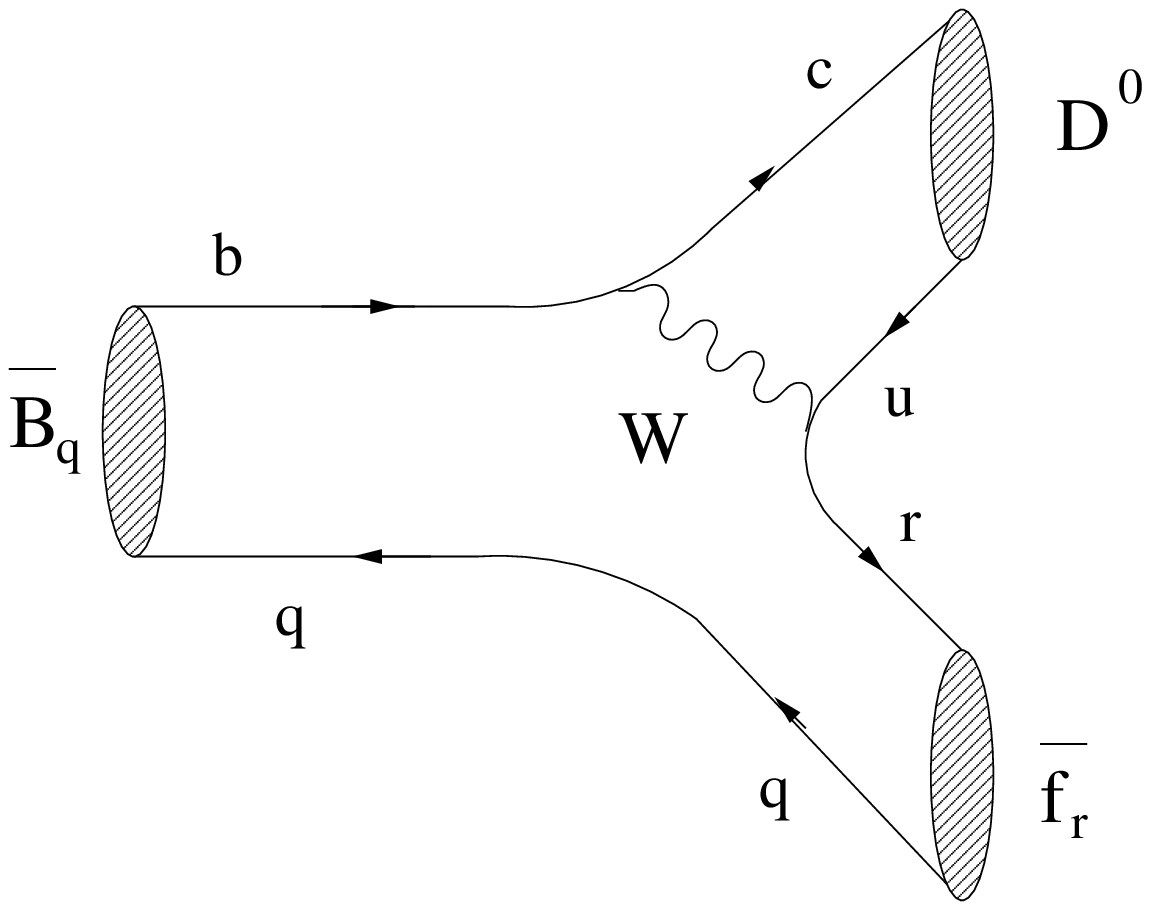}
\end{center}
\caption{Feynman diagrams contributing to $B_q^0\to D^0 f_r$ and 
$\overline{B_q^0}\to D^0\overline{f_r}$.}\label{fig:BqDf}
\end{figure}

\boldmath
\section{Notation, Time Evolution and Decay Amplitudes}\label{sec:Evol-Ampl}
\setcounter{equation}{0}
\unboldmath
\subsection{Notation}\label{subsec:not}
The decays we shall consider in the following analysis may be written 
generically as $B^0_q\to D^0f_r$, $\overline{B^0_q}\to D^0\overline{f_r}$, 
where $q\in\{d,s\}$ and $r\in\{d,s\}$, as can be seen in Fig.~\ref{fig:BqDf}.
Note that the $q$ quark is contained in $f_r$; in order not to complicate 
the notation unnecessarily, we do not indicate this explicitly through 
another label $q$ that should, in principle, be added to $f_r$. If we 
require that $f_r$ be a CP-self-conjugate state, i.e.\ satisfies
\begin{equation}
({\cal CP})|f_r\rangle=\eta_{\rm CP}^{f_r}|f_r\rangle,
\end{equation}
$B^0_q$ and $\overline{B^0_q}$ may both decay into the same state
$D^0f_r$, thereby leading to interference effects between 
$B^0_q$--$\overline{B^0_q}$ mixing and decay processes. Let us illustrate
this abstract but quite efficient notation by giving a few examples:
\begin{itemize}
\item $r=s$: \, $B_d\to DK_{\rm S(L)}$ $(q=d)$; \,
$B_s\to D\eta^{(')}, D\phi, ...$\ $(q=s)$.
\item $r=d$: \, 
$B_s\to DK_{\rm S(L)}$ $(q=s)$; \, $B_d\to D\pi^0, D\rho^0, ...$\ $(q=d)$.
\end{itemize}
All these decays are governed by the colour-suppressed tree-diagram-like 
topologies shown in Fig.~\ref{fig:BqDf}. In the case of $r=q$, also exchange 
topologies may contribute. However, these contributions do not modify the 
phase structure of the corresponding decay amplitudes discussed in 
Subsection~\ref{subsec:ampl}, since they enter with the same CKM factors as 
the colour-suppressed tree topologies. The $\overline{B^0_d}\to D^0\pi^0$ 
mode has already been observed by the Belle, CLEO and BaBar collaborations, 
with branching ratios at the $3\times 10^{-4}$ level \cite{Bbar-D0pi0}.
Interestingly, also the observation of the 
$\overline{B^0_d}\to D^0 \overline{K^0}$ channel was very recently 
announced by the Belle collaboration \cite{Belle-BdDK-obs}, measuring
the branching ratio $(5.0^{+1.3}_{-1.2}\pm0.6)\times10^{-5}$.

\subsection{Time Evolution}\label{time-evol}
If both a $B^0_q$ and a $\overline{B^0_q}$ meson may decay into the
same final state $f$, as is the case for the modes specified above,
we obtain the following time-dependent rate asymmetry \cite{RF-PHYS-REP}:
\begin{eqnarray}
\lefteqn{\frac{\Gamma(B^0_q(t)\to f)-
\Gamma(\overline{B^0_q}(t)\to f)}{\Gamma(B^0_q(t)\to f)+
\Gamma(\overline{B^0_q}(t)\to f)}}\nonumber\\
&&=\left[\frac{C(B_q\to f)\cos(\Delta M_q t)+
S(B_q\to f)\sin(\Delta M_q t)}{\cosh(\Delta\Gamma_qt/2)-{\cal A}_{\rm 
\Delta\Gamma}(B_q\to f)\sinh(\Delta\Gamma_qt/2)}\right],\label{ee6}
\end{eqnarray}
where $\Gamma(B^0_q(t)\to f)$ and $\Gamma(\overline{B^0_q}(t)\to f)$
denote the time-dependent decay rates for initially, i.e.\ at time
$t=0$, present $B^0_q$ and $\overline{B^0_q}$ states, and 
$\Delta M_q\equiv M_{\rm H}^{(q)}-M_{\rm L}^{(q)}>0$ and
$\Delta\Gamma_q\equiv\Gamma_{\rm H}^{(q)}-\Gamma_{\rm L}^{(q)}$
are the mass and decay width differences of the $B_q$ mass eigenstates,
respectively. The observables $C(B_q\to f)$ and $S(B_q\to f)$ are
given by
\begin{equation}\label{obs-expr}
C(B_q\to f)=\frac{1-\bigl|\xi_f^{(q)}\bigr|^2}{1+\bigl|\xi_f^{(q)}\bigr|^2}
\quad \mbox{and} \quad
S(B_q\to f)=\frac{2\,\mbox{Im}\, \xi^{(q)}_f}{1+\bigl|\xi^{(q)}_f\bigr|^2},
\end{equation}
respectively, where
\begin{equation}\label{xi-def}
\xi_f^{(q)}=e^{-i\Theta_{{M}_{12}}^{(q)}}
\frac{A(\overline{B_q^0}\to f)}{A(B_q^0\to f)}
\end{equation}
measures the strength of the interference effects between the
$B^0_q$--$\overline{B^0_q}$ mixing and decay processes. Here
\begin{equation}\label{phi-q-def}
\Theta_{{M}_{12}}^{(q)}-\pi+\phi_{\rm CP}(B_q)=
2\,\mbox{arg}(V_{tq}^\ast V_{tb})
\equiv \phi_q\stackrel{\rm SM}{=}\left\{\begin{array}{cl}
+2\beta={\cal O}(50^\circ) & ~~\mbox{($q=d$)}\\
-2\lambda^2\eta={\cal O}(-2^\circ) & 
~~\mbox{($q=s$)}
\end{array}\right.
\end{equation}
is the CP-violating weak $B^0_q$--$\overline{B^0_q}$ mixing phase, which
is -- within the Standard Model -- negligibly small in the $B_s$-meson
system; the convention-dependent phase $\phi_{\rm CP}(B_q)$ is defined
through
\begin{equation}\label{B-phase}
({\cal CP})|B^0_q\rangle=e^{i\phi_{\rm CP}(B_q)}|\overline{B^0_q}\rangle,
\quad
({\cal CP})|\overline{B^0_q}\rangle=e^{-i\phi_{\rm CP}(B_q)}|B^0_q\rangle.
\end{equation}
As we will see below, $\phi_{\rm CP}(B_q)$ is cancelled through the amplitude 
ratio in (\ref{xi-def}). The width difference $\Delta\Gamma_q$ provides 
another observable,
\begin{equation}\label{ADGam}
{\cal A}_{\rm \Delta\Gamma}(B_q\to f)=
\frac{2\,\mbox{Re}\,\xi^{(q)}_f}{1+\bigl|\xi^{(q)}_f
\bigr|^2},
\end{equation}
which is, however, not independent from $C(B_q\to f)$ and $S(B_q\to f)$,
satisfying
\begin{equation}\label{Obs-rel}
\Bigl[C(B_q\to f)\Bigr]^2+
\Bigl[S(B_q\to f)\Bigr]^2+
\Bigl[{\cal A}_{\Delta\Gamma}(B_q\to f)\Bigr]^2=1.
\end{equation}
This observable could be determined from the ``untagged'' rate
\begin{eqnarray}
\langle\Gamma(B_q(t)\to f)\rangle&\equiv&
\Gamma(B^0_q(t)\to f)+\Gamma(\overline{B^0_q}(t)\to f)=
\left[\Gamma(B^0_q\to f)+\Gamma(\overline{B^0_q}\to f)\right]\nonumber\\
&&\times\left[\cosh(\Delta\Gamma_qt/2)-{\cal A}_{\rm \Delta\Gamma}(B_q\to f)
\,\sinh(\Delta\Gamma_qt/2)\right]e^{-\Gamma_qt},\label{untagged}
\end{eqnarray}
where the oscillatory $\cos(\Delta M_qt)$ and $\sin(\Delta M_qt)$ terms 
cancel, and $\Gamma_q\equiv(\Gamma_{\rm H}^{(q)}+\Gamma_{\rm L}^{(q)})/2$ 
denotes the average decay width \cite{dunietz}. However, in the case of the 
$B_d$-meson system, the width difference is negligibly small, so that the 
time evolution of (\ref{untagged}) is essentially given by the well-known 
exponential $e^{-\Gamma_dt}$. On the other hand, the width difference 
$\Delta \Gamma_s$ of the $B_s$-meson system may be as large as 
${\cal O}(-10\%)$ (for a recent review, see \cite{BeLe}). 

In the discussion given below, we shall see -- as in \cite{RF-BD-CP} -- that 
observables provided by the untagged rates specified in (\ref{untagged}) are 
a key element for an efficient determination of $\gamma$ from $B_q\to Df_s$ 
modes. In contrast to the strategies proposed in \cite{dunietz,DF}, we do 
{\it not} have to rely on a sizeable width difference $\Delta\Gamma_q$; the 
untagged measurements are only required to extract the ``unevolved'', 
i.e.\ time-independent, untagged rates 
\begin{equation}\label{un-tag-rate1}
\langle\Gamma(B_q\to f)\rangle\equiv\Gamma(B^0_q\to f)+
\Gamma(\overline{B^0_q}\to f).
\end{equation}
Consequently, we will not discuss the effects that may be caused by
a possible sizeable value of $\Delta\Gamma_s$ in further detail. They 
could be taken into account straightforwardly.

Concerning the time evolution of the $B_q\to D f_r$ modes analysed in
this paper, the subsequent decay of the neutral $D$-meson state plays a 
crucial r\^ole. Let us assume that this decay, $D\to f_D$, occurs at time 
$t=t'$ after the decay of the $B_q$ meson into $D f_r$. In this case, we
have to consider decay amplitudes of the kind
\begin{equation}\label{A-B-D-evol}
A(B^0_q(t)\to D(t')[\to f_D]f_r),
\end{equation}
and have in principle to deal with $D^0$--$\overline{D^0}$ mixing for $t'>0$.
However, within the Standard Model, this phenomenon is very small. 
Neglecting hence the mixing effects, the time evolution of the neutral 
$|D\rangle$ states is given by
\begin{equation}
|D^0(t')\rangle=e^{-iM_D t'}e^{-\Gamma_Dt'/2}|D^0\rangle, \quad
|\overline{D^0}(t')\rangle=e^{-iM_D t'}e^{-\Gamma_Dt'/2}
|\overline{D^0}\rangle,
\end{equation}
so that (\ref{A-B-D-evol}) yields a time-dependent rate of the following
simple structure:
\begin{equation}\label{Gam-B-D-evol}
\Gamma(B^0_q(t)\to D(t')[\to f_D]f_r)=e^{-\Gamma_Dt'}
\Gamma(B^0_q(t)\to D[\to f_D]f_r),
\end{equation}
where $\Gamma(B^0_q(t)\to D[\to f_D]f_r)$ corresponds to the 
rates entering (\ref{ee6}), and $\Gamma_D$ denotes the decay width of the
neutral $D$ mesons. As is well known, $D^0$--$\overline{D^0}$ mixing may 
be enhanced through new physics. Should this actually be the case, the 
mixing effects could be included through a measurement of the 
$D^0$--$\overline{D^0}$ mixing parameters \cite{D-studies}. Note that 
$D^0$--$\overline{D^0}$ mixing does not lead to any effects in the case 
of $t'=0$, even if this phenomenon should be made sizeable by the presence 
of new physics. In the following considerations, we shall neglect 
$D^0$--$\overline{D^0}$ mixing for simplicity. Moreover, we shall 
assume that CP violation in $D$ decays is also negligible, as in the 
Standard Model. In analogy to $D^0$--$\overline{D^0}$ mixing, these effects 
could also be enhanced through new physics. In the following sections, 
we will therefore emphasize clearly whenever this assumption enters. 

Note that also the tiny ``indirect'' CP violation in the neutral kaon system 
has -- at least in principle -- to be taken into account if $K_{\rm S}$ or 
$K_{\rm L}$ states are involved. This could be done through the corresponding 
experimental $K$-decay data. However, it appears very unlikely that the 
experimental accuracy of the strategies for the determination of $\gamma$
discussed below will ever reach a level where these effects may play any 
r\^ole from a practical point of view.

\subsection{Decay Amplitudes}\label{subsec:ampl}
In order to analyse the structure of the $B_q\to Df_r$ decay amplitudes, 
we employ appropriate low-energy effective Hamiltonians \cite{RF-PHYS-REP}:
\begin{eqnarray}
{\cal H}_{\rm eff}(\overline{B^0_q}\to D^0f_r)&=&
\frac{G_{\rm F}}{\sqrt{2}}\,\overline{v}_r\left[
\overline{{\cal O}}_1^{\, r}\,{ C}_1(\mu)+
\overline{{\cal O}}_2^{\, r}\,{ C}_2(\mu)\right]\label{Heff-bar}\\
{\cal H}_{\mbox{{\scriptsize eff}}}(B^0_q\to D^0f_r)&=&
\frac{G_{\rm F}}{\sqrt{2}}\,v_r^\ast
\left[{\cal O}_1^{r\dagger}\,{ C}_1(\mu)+{\cal O}_2^{r\dagger}\,
{ C}_2(\mu)\right],\label{Heff}
\end{eqnarray}
where $\overline{{\cal O}}_k^{\, r}$ and ${\cal O}_k^{\, r}$ denote 
current--current operators, which are given by
\begin{equation}
\begin{array}{rclrcl}
\overline{{\cal O}}_1^{\, r}
&=&(\overline{r}_\alpha u_\beta)_{\mbox{{\scriptsize 
V--A}}}\left(\overline{c}_\beta b_\alpha\right)_{\mbox{{\scriptsize V--A}}},&
~~\overline{{\cal O}}_2^{\, r}
&=&(\overline{r}_\alpha u_\alpha)_{\mbox{{\scriptsize 
V--A}}}\left(\overline{c}_\beta b_\beta\right)_{\mbox{{\scriptsize V--A}}},\\
&&&&&\\
{\cal O}_1^{\, r}&=&(\overline{r}_\alpha c_\beta)_{\mbox{{\scriptsize V--A}}}
\left(\overline{u}_\beta b_\alpha\right)_{\mbox{{\scriptsize V--A}}},&
~~{\cal O}_2^{\, r}
&=&(\overline{r}_\alpha c_\alpha)_{\mbox{{\scriptsize V--A}}}
\left(\overline{u}_\beta b_\beta\right)_{\mbox{{\scriptsize V--A}}},
\end{array}
\end{equation}
and 
\begin{equation}\label{CKM-fact1}
v_s=V_{cs}^\ast V_{ub}=
A\lambda^3R_be^{-i\gamma}, \quad 
\overline{v}_s=V_{us}^\ast V_{cb}=A\lambda^3,
\end{equation}
\begin{equation}\label{CKM-fact2}
v_d=V_{cd}^\ast V_{ub}=
-\left(\frac{A\lambda^4R_b}{1-\lambda^2/2}\right)e^{-i\gamma},  \quad 
\overline{v}_d=V_{ud}^\ast V_{cb}=A\lambda^2(1-\lambda^2/2)
\end{equation}
are CKM factors, with (for the numerical values, see \cite{Andr-02})
\begin{equation}
A\equiv\frac{1}{\lambda^2}|V_{cb}|=0.83\pm0.02,
\end{equation}
\begin{equation}
R_b\equiv\left(1-\frac{\lambda^2}{2}\right)\frac{1}{\lambda}\left|
\frac{V_{ub}}{V_{cb}}\right|=\sqrt{\overline{\rho}^2+\overline{\eta}^2}
=0.39\pm0.04.
\end{equation}
Using (\ref{Heff-bar}), we may write
\begin{equation}
A(\overline{B^0_q}\to D^0f_r)=\langle f_rD^0|
{\cal H}_{\rm eff}(\overline{B^0_q}\to D^0f_r)|\overline{B^0_q}\rangle=
\frac{G_{\rm F}}{\sqrt{2}}\overline{v}_r\overline{M}_{f_r},
\end{equation}
where
\begin{equation}
\overline{M}_{f_r}\equiv 
\langle f_rD^0|\overline{{\cal O}}_1^{\, r}\,{ C}_1(\mu)+
\overline{{\cal O}}_2^{\, r}\,{ C}_2(\mu)|\overline{B^0_q}\rangle.
\end{equation}
On the other hand, we have
\begin{equation}
A(B^0_q\to D^0f_r)=\langle f_rD^0|{\cal H}_{\rm eff}(B^0_q\to D^0f_r)
|B^0_q\rangle,
\end{equation}
which we may rewrite as 
\begin{eqnarray}
A(B^0_q\to D^0f_r)&=&\langle f_rD^0|({\cal CP})^\dagger ({\cal CP})
{\cal H}_{\rm eff}(B^0_q\to D^0f_r)({\cal CP})^\dagger 
({\cal CP})|B^0_q\rangle\nonumber\\
&=&(-1)^L \eta_{\rm CP}^{f_r}e^{i[\phi_{\rm CP}(B_q)-\phi_{\rm CP}(D)]}
\frac{G_{\rm F}}{\sqrt{2}}\,v_r^\ast M_{f_r},
\end{eqnarray}
where
\begin{equation}
M_{f_r}\equiv \langle f_r\overline{D^0}|{\cal O}_1^{\, r}\,{ C}_1(\mu)+
{\cal O}_2^{\, r}\,{ C}_2(\mu)|\overline{B^0_q}\rangle.
\end{equation}
Here we have employed
\begin{equation}
({\cal CP})^\dagger ({\cal CP})=\hat 1,
\end{equation}
\begin{equation}
({\cal CP}){\cal O}_k^{\, r\dagger}({\cal CP})^\dagger=
{\cal O}_k^{\, r},
\end{equation}
as well as
\begin{equation}
({\cal CP})|D^0f_r\rangle=(-1)^L\eta_{\rm CP}^{f_r}
e^{i\phi_{\rm CP}(D)}|\overline{D^0}f_r\rangle,
\end{equation}
where $L$ denotes the angular momentum of the $D^0f_r$ state, and 
\begin{equation}\label{D-phase}
({\cal CP})|D^0\rangle=e^{i\phi_{\rm CP}(D)}|\overline{D^0}\rangle,
\quad
({\cal CP})|\overline{D^0}\rangle=e^{-i\phi_{\rm CP}(D)}|D^0\rangle,
\end{equation}
in analogy to (\ref{B-phase}). In the following, we shall use 
the abbreviation 
\begin{equation}
\eta_{f_r}\equiv (-1)^L\eta_{\rm CP}^{f_r}.
\end{equation}
For convenience, we have collected in Table~\ref{tab:CP} the values of $L$, 
$\eta_{\rm CP}^{f_r}$ and $\eta_{f_r}$ for various decay channels of the 
kind specified in Subsection~\ref{subsec:not}.

\begin{table}
\begin{center}
\begin{tabular}{|l|l||l|l|l|}
\hline
~~~~~~$r=s$ & ~~~~~~$r=d$ & ~~$L$~~ & ~~$\eta_{\rm CP}^{f_r}$~~ & 
~~$\eta_{f_r}$~~ \\
\hline\hline
$\overline{B^0_d}\to D^0 K_{\rm S (L)}$ & 
$\overline{B^0_s}\to D^0 K_{\rm S (L)}$ &
~~0~~ & ~$+1 (-1)$~ & ~$+1 (-1)$~\\
$\overline{B^0_s}\to D^0 \eta^{(')}$ & 
$\overline{B^0_d}\to D^0 \pi^0$ & ~~0~~ & ~$-1$~ & ~$-1$~\\
$\overline{B^0_s}\to D^0 \phi$ & 
$\overline{B^0_d}\to D^0 \rho^0$ & ~~1~~ & ~$+1$~ & ~$-1$~\\
\hline
\end{tabular}
\caption{The values of $L$, $\eta_{\rm CP}^{f_r}$ and $\eta_{f_r}$ for
various $\overline{B_q^0}\to D^0 f_r$ modes.}\label{tab:CP}
\end{center}
\end{table}

Let us now consider the decays $\overline{B^0_q}\to \overline{D^0}f_r$
and  $B^0_q\to \overline{D^0}f_r$, which are described by 
\begin{eqnarray}
{\cal H}_{\mbox{{\scriptsize eff}}}(\overline{B^0_q}\to\overline{D^0}f_r)&=&
\frac{G_{\rm F}}{\sqrt{2}}\,v_r\left[{\cal O}_1^r\,{ C}_1(\mu)+
{\cal O}_2^r\,{ C}_2(\mu)\right]\label{Heff2}\\
{\cal H}_{\rm eff}(B^0_q\to\overline{D^0}f_r)&=&
\frac{G_{\rm F}}{\sqrt{2}}\,\overline{v}_r^\ast\left[
\overline{{\cal O}}_1^{\, r\dagger}\,{ C}_1(\mu)+
\overline{{\cal O}}_2^{\, r\dagger}\,{ C}_2(\mu)\right],\label{Heff-bar2}
\end{eqnarray}
yielding the following transition amplitudes:
\begin{equation}
A(\overline{B^0_q}\to \overline{D^0}f_r)=\frac{G_{\rm F}}{\sqrt{2}}v_rM_{f_r}
\end{equation}
\begin{equation}
A(B^0_q\to\overline{D^0}f_r)=\eta_{f_r} 
e^{i[\phi_{\rm CP}(B_q)+\phi_{\rm CP}(D)]}\frac{G_{\rm F}}{\sqrt{2}}
\overline{v}_r^\ast
\overline{M}_{f_r}.
\end{equation}

For the time-dependent decay rates, the following amplitude ratios play
a key r\^ole:
\begin{equation}\label{rat1}
\frac{A(\overline{B^0_q}\to D^0f_r)}{A(B^0_q\to D^0f_r)}
=\eta_{f_r} e^{-i\phi_{\rm CP}(B_q)}
\frac{\overline{v}_r}{v_r^\ast}\frac{1}{a_{f_r}e^{i\delta_{f_r}}}
\end{equation}
\begin{equation}\label{rat2}
\frac{A(\overline{B^0_q}\to \overline{D^0}f_r)}{A(B^0_q\to 
\overline{D^0}f_r)}
=\eta_{f_r} e^{-i\phi_{\rm CP}(B_q)}
\frac{v_r}{\overline{v}_r^\ast}a_{f_r}e^{i\delta_{f_r}},
\end{equation}
where
\begin{equation}\label{a-def}
a_{f_r}e^{i\delta_{f_r}}\equiv e^{-i\phi_{\rm CP}(D)}
\frac{M_{f_r}}{\overline{M}_{f_r}}=e^{-i\phi_{\rm CP}(D)}
\left[\frac{\langle f_r\overline{D^0}|
{\cal O}_1^{\, r}\,{ C}_1(\mu)+{\cal O}_2^{\, r}\,{ C}_2(\mu)|
\overline{B^0_q}\rangle}{\langle f_rD^0|
\overline{{\cal O}}_1^{\, r}\,{ C}_1(\mu)+
\overline{{\cal O}}_2^{\, r}\,{ C}_2(\mu)|
\overline{B^0_q}\rangle}\right].
\end{equation}
Let us illustrate the calculation of the hadronic parameter 
$a_{f_r}e^{i\delta_{f_r}}$ using the factorization approach,
which yields
\begin{displaymath}
\left.\langle f_r\overline{D^0}|{\cal O}_1^{\, r}\,{ C}_1(\mu)+
{\cal O}_2^{\, r}\,{ C}_2(\mu)|\overline{B^0_q}\rangle\right|_{\rm fact}
=a_2\langle\overline{D^0}|(\overline{u}_\beta c_\beta)_{\mbox{{\scriptsize 
V--A}}}|0\rangle\langle f_r|(\overline{r}_\alpha b_\alpha)_{\mbox{{\scriptsize 
V--A}}}|\overline{B^0_q}\rangle
\end{displaymath}
\begin{equation}\label{ME-fact1}
=-e^{i\phi_{\rm CP}(D)}
a_2\langle D^0|(\overline{c}_\beta u_\beta)_{\mbox{{\scriptsize 
V--A}}}|0\rangle\langle f_r|(\overline{r}_\alpha b_\alpha)_{\mbox{{\scriptsize 
V--A}}}|\overline{B^0_q}\rangle
\end{equation}
and
\begin{equation}\label{ME-fact2}
\left.\langle f_r D^0|\overline{{\cal O}}_1^{\, r}\,{ C}_1(\mu)+
\overline{{\cal O}}_2^{\, r}\,{ C}_2(\mu)|\overline{B^0_q}\rangle
\right|_{\rm fact}
=a_2\langle D^0|(\overline{c}_\beta u_\beta)_{\mbox{{\scriptsize 
V--A}}}|0\rangle\langle f_r|(\overline{r}_\alpha b_\alpha)_{\mbox{{\scriptsize 
V--A}}}|\overline{B^0_q}\rangle,
\end{equation}
where
\begin{equation}
a_2=C_1(\mu_{\rm F})+\frac{C_2(\mu_{\rm F})}{N_{\rm C}}
\end{equation}
is the well-known colour-suppression factor, with a factorization
scale $\mu_{\rm F}$ and a number $N_{\rm C}$ of quark colours. 
Inserting (\ref{ME-fact1}) and (\ref{ME-fact2}) into (\ref{a-def}),
we observe that the convention-dependent phase $\phi_{\rm CP}(D)$ 
cancels -- as it has to -- and eventually arrive at the simple
result
\begin{equation}\label{a-fact}
\left.a_{f_r}e^{i\delta_{f_r}}\right|_{\rm fact}=-1,
\end{equation}
where the factorized hadronic matrix elements cancel as well. In particular,
we conclude that $\left.\delta_{f_r}\right|_{\rm fact}=180^\circ$. Here it 
pays off to have kept the convention-dependent phase $\phi_{\rm CP}(D)$ 
explicitly in our calculation. If we look at the Feynman diagrams shown in
Fig.~\ref{fig:BqDf} and contract the propagators of the $W$ bosons, we
observe that the hadronic matrix elements entering (\ref{a-def}) are related
to each other through an interchange of all up and charm quarks and are
hence very similar. Since $a_{f_r}e^{i\delta_{f_r}}$ is governed by the
{\it ratio} of these matrix elements, and $\delta_{f_r}$ measures the 
{\it relative} strong phase between them, we expect $\delta_{f_r}$ to 
differ moderatly from the trivial value of $180^\circ$, 
even if the hadronic matrix elements themselves may 
deviate sizeably from the factorization case, as advocated in 
\cite{non-fact} (note that the recently developed QCD factorization approach
\cite{QCD-fact-tree} does not apply to $B_q\to D f_r$ modes). In particular, 
the assumption 
\begin{equation}\label{dyn-assumpt}
\cos\delta_{f_r}<0,
\end{equation}
which is satisfied for the whole range of $90^\circ<\delta_{f_r}<270^\circ$, 
is very plausible.

For the following considerations, it is convenient to
rewrite (\ref{rat1}) and (\ref{rat2}) with the help of (\ref{CKM-fact1})
and (\ref{CKM-fact2}) as follows:
\begin{equation}\label{rat1x}
\frac{A(\overline{B^0_q}\to D^0f_r)}{A(B^0_q\to D^0f_r)}
=\eta_{f_r}e^{-i\phi_{\rm CP}(B_q)}e^{-i\gamma}
\frac{1}{x_{f_r}e^{i\delta_{f_r}}}
\end{equation}
\begin{equation}\label{rat2x}
\frac{A(\overline{B^0_q}\to \overline{D^0}f_r)}{A(B^0_q\to 
\overline{D^0}f_r)}
=\eta_{f_r}e^{-i\phi_{\rm CP}(B_q)}e^{-i\gamma}
x_{f_r}e^{i\delta_{f_r}},
\end{equation}
where
\begin{equation}\label{X-def}
x_{f_s}\equiv R_b a_{f_s}, \quad 
x_{f_d}\equiv-\left(\frac{\lambda^2R_b}{1-\lambda^2}\right)a_{f_d}.
\end{equation}
Note that 
\begin{equation}
\frac{x_{f_d}}{x_{f_s}}\approx-\left(\frac{\lambda^2}{1-\lambda^2}
\right)=-0.05
\end{equation}
reflects that the interference effects between the $\overline{B^0_q}\to
\overline{D^0}f_d$ and $B^0_q\to \overline{D^0}f_d$ decay paths are
doubly Cabibbo-suppressed, in contrast to the favourable situation 
in $\overline{B^0_q}\to \overline{D^0}f_s$ and $B^0_q\to \overline{D^0}f_s$,
showing large interference effects of order $R_b\approx0.4$.

\boldmath
\section{Case of $B_q\to D[\to f_\pm] f_r$}\label{sec:CP}
\setcounter{equation}{0}
\unboldmath
\subsection{Amplitudes}
Let us now apply the formulae derived in the previous section to 
analyse the case where the neutral $D$ mesons are observed through their
decays into CP eigenstates $f_+$ and $f_-$ with eigenvalues $+1$ 
and $-1$, respectively. Important examples are $f_+=K^+K^-,\pi^+\pi^-$ and
$f_-=K_{\rm S}\pi^0, K_{\rm S}\phi$. The 
$\overline{B^0_q}\to D[\to f_\pm]f_r$ amplitude is given by
\begin{displaymath}
A(\overline{B^0_q}\to D[\to f_\pm]f_r)=
A(\overline{B^0_q}\to D^0f_r)A(D^0\to f_\pm)+
A(\overline{B^0_q}\to\overline{D^0}f_r)A(\overline{D^0}\to f_\pm)
\end{displaymath}
\begin{equation}
=\frac{G_{\rm F}}{\sqrt{2}}\overline{v}_r\overline{M}_{f_r}
A(D^0\to f_\pm)\left[1+\frac{v_r}{\overline{v}_r}
\frac{M_{f_r}}{\overline{M}_{f_r}}
\frac{A(\overline{D^0}\to f_\pm)}{A(D^0\to f_\pm)}\right].
\end{equation}
If we neglect CP-violating effects in neutral $D$-meson decays, i.e.\
assume that the corresponding low-energy effective Hamiltonian
${\cal H}_{\rm eff}^{(D)}$ satisfies
\begin{equation}\label{CP-D}
({\cal CP}){\cal H}_{\rm eff}^{(D)}({\cal CP})^\dagger=
{\cal H}_{\rm eff}^{(D)\dagger},
\end{equation}
we have
\begin{displaymath}
A(\overline{D^0}\to f_\pm)=\langle f_\pm|{\cal H}_{\rm eff}^{(D)}
|\overline{D^0}\rangle=\langle f_\pm|({\cal CP})^\dagger
({\cal CP}){\cal H}_{\rm eff}^{(D)}({\cal CP})^\dagger({\cal CP})
|\overline{D^0}\rangle
\end{displaymath}
\begin{equation}
=\pm e^{-i\phi_{\rm CP}(D)}
\langle f_\pm|{\cal H}_{\rm eff}^{(D)\dagger}|D^0\rangle
=\pm e^{-i\phi_{\rm CP}(D)}A(D^0\to f_\pm).
\end{equation}
In fact, (\ref{CP-D}) is fulfilled to excellent accuracy in the
Standard Model (see the comment at the end of Subsection~\ref{time-evol}). 
Consequently, we obtain
\begin{equation}
A(\overline{B^0_q}\to D[\to f_\pm]f_r)
=\frac{G_{\rm F}}{\sqrt{2}}\overline{v}_r\overline{M}_{f_r}
A(D^0\to f_\pm)\left[1\pm\frac{v_r}{\overline{v}_r}
a_{f_r}e^{i\delta_{f_r}}\right].
\end{equation}
Performing an analogous calculation for the $B^0_q\to D[\to f_\pm]f_r$
case yields
\begin{equation}
A(B^0_q\to D[\to f_\pm]f_r)=\pm
\eta_{f_r} e^{i\phi_{\rm CP}(B_q)}\frac{G_{\rm F}}{\sqrt{2}}
\overline{v}_r^\ast\overline{M}_{f_r}A(D^0\to f_\pm)
\left[1\pm\left(\frac{v_r}{\overline{v}_r}\right)^\ast
a_{f_r}e^{i\delta_{f_r}}\right].
\end{equation}

\boldmath
\subsection{Observables of the Time-Dependent Decay Rates}
\unboldmath
If we take into account (\ref{CKM-fact1}), (\ref{CKM-fact2}) and 
(\ref{X-def}), we obtain the following expression for observable 
$\xi_{D[\to f_\pm]f_r}^{(q)}$ introduced in (\ref{xi-def}):
\begin{equation}
\xi_{D[\to f_\pm]f_r}^{(q)}\equiv\xi_\pm^{(q)}=
\mp \eta_{f_r} e^{-i\phi_q}
\left[\frac{1\pm e^{-i\gamma}x_{f_r}e^{i\delta_{f_r}}}{1\pm 
e^{+i\gamma}x_{f_r}e^{i\delta_{f_r}}}\right],
\end{equation}
where $\Theta_{M_{12}}^{(q)}$ cancels the phase $\phi_{\rm CP}(B_q)$
appearing in the amplitude ratio, thereby yielding a convention-independent 
result.  Employing (\ref{obs-expr}), we arrive at the following expressions 
for the observables provided by the time-dependent decay rates:
\begin{equation}
C(B_q\to D[\to f_\pm]f_r)\equiv C_\pm^{f_r}=
\mp\left[\frac{2\,x_{f_r}\sin\delta_{f_r}\sin\gamma}{1\pm
2\,x_{f_r}\cos\delta_{f_r}\cos\gamma+x_{f_r}^2}\right],
\end{equation}
\begin{eqnarray}
\lefteqn{S(B_q\to D[\to f_\pm]f_r)\equiv S_\pm^{f_r}}\nonumber\\
&&=\pm\,\eta_{f_r}
\left[\frac{\sin\phi_q \pm 2\,x_{f_r}\cos\delta_{f_r}\sin(\phi_q+\gamma)+
x_{f_r}^2\sin(\phi_q+2\gamma)}{1 \pm 2\,x_{f_r}\cos\delta_{f_r}\cos\gamma
+x_{f_r}^2}\right].
\end{eqnarray}
As noted in \cite{RF-BD-CP}, it is convenient to consider the following 
combinations:
\begin{equation}\label{Cplus-def}
\langle C_{f_r}\rangle_+\equiv\frac{C_+^{f_r}+C_-^{f_r}}{2}
=\frac{x_{f_r}^2\sin2\delta_{f_r}\sin2\gamma}{(1+x_{f_r}^2)^2-(2\,x_{f_r}
\cos\delta_{f_r}\cos\gamma)^2}
\end{equation}
\begin{equation}\label{Cminus-def}
\langle C_{f_r}\rangle_-\equiv\frac{C_+^{f_r}-C_-^{f_r}}{2}
=-\left[\frac{2\,x_{f_r}(1+x_{f_r}^2)\sin\delta_{f_r}
\sin\gamma}{(1+x_{f_r}^2)^2-(2\,x_{f_r}\cos\delta_{f_r}\cos\gamma)^2}\right]
\end{equation}
\begin{equation}\label{Splus-CP}
\langle S_{f_r}\rangle_+\equiv\frac{S_+^{f_r}+S_-^{f_r}}{2}
=\eta_{f_r}\left[\frac{2\,x_{f_r}\cos\delta_{f_r}\sin\gamma
\left\{\cos\phi_q-x_{f_r}^2\cos(\phi_q+2\gamma)\right\}}{(1+x_{f_r}^2)^2-
(2\,x_{f_r}\cos\delta_{f_r}\cos\gamma)^2}\right]
\end{equation}
\begin{eqnarray}
\lefteqn{\langle S_{f_r}\rangle_-\equiv\frac{S_+^{f_r}-S_-^{f_r}}{2}}
\label{Sminus-CP}\\
&&=\eta_{f_r}\left[\frac{\sin\phi_q+x_{f_r}^2\left\{\sin\phi_q+
(1+x_{f_r}^2)\sin(\phi_q+2\gamma)-4\cos^2\delta_{f_r}
\cos\gamma\sin(\phi_q+\gamma)\right\}}{(1+x_{f_r}^2)^2-
(2\,x_{f_r}\cos\delta_{f_r}\cos\gamma)^2}\right].\nonumber
\end{eqnarray}
If we use
\begin{equation}\label{Gpm-def}
-\left[\frac{\langle C_{f_r}\rangle_+}{\langle C_{f_r}\rangle_-}\right]=
\frac{2\,x_{f_r}\cos\delta_{f_r}\cos\gamma}{1+x_{f_r}^2}\equiv
\Gamma_{+-}^{f_r},
\end{equation}
we may write (\ref{Splus-CP}) as 
\begin{equation}
\eta_{f_r} \langle S_{f_r}\rangle_+=\left[\frac{\Gamma_{+-}^{f_r}
\tan\gamma}{1-(\Gamma_{+-}^{f_r})^2}\right]
\left[\cos\phi_q-\frac{2\,x_{f_r}^2}{1+x_{f_r}^2}\cos(\phi_q+\gamma)
\cos\gamma\right].
\end{equation}
Employing, in addition, expression (\ref{Sminus-CP}) for
$\langle S_{f_r}\rangle_-$, we eventually arrive at the
following transparent final result \cite{RF-BD-CP}:
\begin{equation}\label{tan-gam-CP}
\tan\gamma\cos\phi_q=
\left[\frac{\eta_{f_r} \langle S_{f_r}\rangle_+}{\Gamma_{+-}^{f_r}}\right]
+\left[\eta_{f_r}\langle S_{f_r}\rangle_--\sin\phi_q\right],
\end{equation}
where the first term in square brackets is ${\cal O}(1)$ and the second one 
involving $\eta_{f_r}\langle S_{f_r}\rangle_-$ is of order $x_{f_r}^2$,
playing a minor r\^ole.  It is important to emphasize that (\ref{tan-gam-CP}) 
is an {\it exact} relation. In particular, it does not rely on any 
assumptions related to factorization or the strong phase $\delta_{f_r}$. 
Since the $B^0_q$--$\overline{B^0_q}$ mixing phase $\phi_q$ can be determined
separately \cite{RF-PHYS-REP}, (\ref{tan-gam-CP}) allows a 
{\it theoretically clean} extraction of $\gamma$.

\boldmath
\subsection{Untagged Observables}
\unboldmath
As pointed out in \cite{RF-BD-CP}, an interesting avenue to determine the
quantity $\Gamma_{+-}^{f_r}$ is provided by the untagged rates specified
in (\ref{untagged}). In comparison with the extraction of 
$\Gamma_{+-}^{f_r}$ from (\ref{Gpm-def}), this method is much more 
promising from a practical point of view. In \cite{RF-BD-CP}, we have 
used the CP-even and CP-odd eigenstates $D_+$ and $D_-$ of the neutral 
$D$-meson system, 
\begin{equation}\label{D-CP-def}
|D_\pm\rangle=\frac{1}{\sqrt{2}}\left[|D^0\rangle\pm e^{i\phi_{\rm CP}(D)}
|\overline{D^0}\rangle\right],
\end{equation}
as a starting point to derive the corresponding formulae. Let us here 
consider explicitly the $B_q\to D[\to f_\pm] f_r$ decay processes, as in the 
previous subsections. Following these lines, we obtain 
\begin{equation}\label{untag1}
\frac{\langle\Gamma(B_q\to D[\to f_\pm]f_r)\rangle}{2\,\Gamma(D^0\to f_\pm)} 
=\Gamma(\overline{B^0_q}\to D^0f_r)
\left[1\pm2\,x_{f_r}\cos\delta_{f_r}\cos\gamma+x_{f_r}^2\right],
\end{equation}
where
\begin{equation}
\langle\Gamma(B_q\to D[\to f_\pm]f_r)\rangle=\Gamma(B_q^0\to D[\to f_\pm]f_r)
+\Gamma(\overline{B_q^0}\to D[\to f_\pm]f_r),
\end{equation}
in accordance with (\ref{un-tag-rate1}). Consequently, if we introduce
\begin{equation}
R_{f_{+/-}}\equiv\frac{\Gamma(D^0\to f_+)}{\Gamma(D^0\to f_-)},
\end{equation}
we obtain
\begin{equation}\label{untag-asym1}
\frac{\langle\Gamma(B_q\to D[\to f_+]f_r)\rangle-
\langle\Gamma(B_q\to D[\to f_-]f_r)\rangle R_{f_{+/-}}}{\langle\Gamma(B_q\to 
D[\to f_+]f_r)\rangle+\langle\Gamma(B_q\to D[\to f_-]f_r)\rangle R_{f_{+/-}}}=
\Gamma_{+-}^{f_r}.
\end{equation}
This relation holds for any given CP eigenstates $f_+$ and $f_-$, and
allows a determination of $\Gamma_{+-}^{f_r}$ from the corresponding
untagged $B_q$-decay data samples. It should be emphasized that the 
unevolved rate $\Gamma(\overline{B^0_q}\to D^0f_r)$ appearing in 
(\ref{untag1}) cancels in (\ref{untag-asym1}). Since the measurement
of this rate through hadronic tags of the $D^0$ meson of the kind
$D^0\to\pi^+K^-$ is affected by certain interference effects \cite{KaLo,ADS}, 
as we will see in Section~\ref{sec:CPNE}, this feature is an important 
advantage. Let us note that also the observables 
(\ref{Cplus-def})--(\ref{Sminus-CP}) provided by the tagged, 
time-dependent $B_q\to D[\to f_\pm]f_r$ decay rates do not involve 
$\Gamma(\overline{B^0_q}\to D^0f_r)$. In contrast to (\ref{Gpm-def}), 
(\ref{untag-asym1}) requires knowledge of $R_{f_{+/-}}$, which has to be 
determined from experimental $D$-decay studies. We have already plenty 
of information on such decays available \cite{PDG}, which will become much 
richer by the time the strategies proposed in this paper can be performed
in practice. 

Let us now turn to the discussion in terms of $D_\pm$ states given in 
\cite{RF-BD-CP}, corresponding to the following rates:
\begin{equation}
\langle\Gamma(B_q\to D_\pm f_r)\rangle=\sum_{f_\pm}
\langle\Gamma(B_q\to D[\to f_\pm]f_r)\rangle,
\end{equation}
which yield
\begin{equation}\label{untag-asym2}
\Gamma_{+-}^{f_r}=\frac{\langle\Gamma(B_q\to D_+f_r)\rangle-
\langle\Gamma(B_q\to D_-f_r)\rangle R_{D_{+/-}}}{\langle\Gamma(B_q\to 
D_+f_r)\rangle+\langle\Gamma(B_q\to D_-f_r)\rangle R_{D_{+/-}}},
\end{equation}
where
\begin{equation}
R_{D_{+/-}}\equiv
\frac{\sum_{f_+}\Gamma(D^0\to f_+)}{\sum_{f_-}\Gamma(D^0\to f_-)}.
\end{equation}
If we compare (\ref{untag-asym2}) with the corresponding expression given 
in \cite{RF-BD-CP}, we observe that the $R_{D_{+/-}}$ factors are missing. 
Indeed, taking into account that 
\begin{equation}
y_{\rm CP}\equiv\frac{\sum_{f_+}\Gamma(D^0\to f_+)-
\sum_{f_-}\Gamma(D^0\to f_-)}{\sum_{f_+}\Gamma(D^0\to f_+)+
\sum_{f_-}\Gamma(D^0\to f_-)}
\end{equation}
is one of the $D^0$--$\overline{D^0}$ mixing parameters \cite{ligeti-D-rev}, 
we find
\begin{equation}
R_{D_{+/-}}=\frac{1+y_{\rm CP}}{1-y_{\rm CP}}=1+2y_{\rm CP}+
{\cal O}(y_{\rm CP}^2).
\end{equation} 
If $D^0$--$\overline{D^0}$ mixing effects are neglected, as done in 
\cite{RF-BD-CP}, we have $R_{D_{+/-}}=1$, yielding
\begin{equation}\label{Gpm-untagged}
\Gamma_{+-}^{f_r}=\frac{\langle\Gamma(B_q\to D_+f_r)\rangle-
\langle\Gamma(B_q\to D_-f_r)\rangle}{\langle\Gamma(B_q\to 
D_+f_r)\rangle+\langle\Gamma(B_q\to D_-f_r)\rangle}=
\frac{2\,x_{f_r}\cos\delta_{f_r}\cos\gamma}{1+x_{f_r}^2}.
\end{equation}
Since the world average for $y_{\rm CP}$ is given by
\cite{ligeti-D-rev}
\begin{equation}
y_{\rm CP}=(1.0\pm0.7)\%,
\end{equation}
the experimental constraint on the deviation of $R_{D_{+/-}}$ from 1 is
already at the per cent level. As pointed out in \cite{RF-BD-CP}, 
(\ref{Gpm-untagged}) offers a ``gold-plated'' way to determine the 
observable $\Gamma_{+-}^{f_r}$, which is a key element for efficient 
determinations of $\gamma$.

\boldmath
\subsection{Applications}\label{subsec:appl-CP}
\unboldmath
The formulae derived in the previous subsections have important
applications for the exploration of CP violation, as pointed out 
and discussed in detail in \cite{RF-BD-CP}:
\begin{itemize}
\item $B_d\to D[\to f_\pm]K_{\rm S(L)}$, $B_s\to D[\to f_\pm]\eta^{(')}$, 
$D[\to f_\pm]\phi$, ..., i.e.\ $r=s$:
using (\ref{Gpm-untagged})
to determine $\Gamma_{+-}^{f_s}$, (\ref{tan-gam-CP}) provides very 
efficient, theoretically clean determinations of $\gamma$ in an essentially 
unambiguous manner. The hadronic parameters $\delta_{f_s}$ and $x_{f_s}$
can be extracted as well, thereby offering interesting insights into
hadronic physics. 
\item $B_s\to D[\to f_\pm]K_{\rm S(L)}$, $B_d\to D[\to f_\pm]\pi^0$, 
$D[\to f_\pm]\rho^0$, ..., i.e.\ $r=d$:
since $x_{f_d}/x_{f_s}\approx-0.05$ is doubly Cabibbo-suppressed, 
decays of this kind do not appear as promising for the extraction of $\gamma$
as their $r=s$ counterparts. However, because of 
\begin{equation}\label{sin-phi-det}
\eta_{f_d}\langle S_{f_d}\rangle_-=\sin\phi_q + {\cal O}(x_{f_d}^2)
=\sin\phi_q + {\cal O}(4\times 10^{-4}),
\end{equation}
they allow interesting determinations of $\sin\phi_q$. In comparison with 
the conventional $B_d\to J/\psi K_{\rm S(L)}$ and $B_s\to J/\psi\phi$ 
methods, these extractions do not suffer from any penguin uncertainties, 
and are theoretically cleaner by one order of magnitude. This feature
is particularly interesting for the $B_s$-meson case.
\end{itemize}
Let us now focus on $B_q\to D[\to f_{\rm NE}] f_r$ transitions, where the
neutral $D$ mesons are observed through decays into CP non-eigenstates
$f_{\rm NE}$. Also in this interesting case, (\ref{Gpm-untagged}) provides
a key ingredient for an efficient and essentially unambiguous determination 
of $\gamma$ from the $r=s$ modes.

\boldmath
\section{Case of $B_q\to D[\to f_{\rm NE}] f_r$}\label{sec:CPNE}
\setcounter{equation}{0}
\unboldmath
Let us now consider decays of the neutral $D$ mesons into CP non-eigenstates 
$f_{\rm NE}$ and their CP conjugates $\overline{f}_{\rm NE}$, which 
satisfy -- in analogy to (\ref{B-phase}) and (\ref{D-phase}) -- the 
following relations:
\begin{equation}
({\cal CP})|f_{\rm NE}\rangle = e^{i\phi_{\rm CP}(f_{\rm NE})}
|\overline{f}_{\rm NE}\rangle, \quad 
({\cal CP})|\overline{f}_{\rm NE}\rangle = e^{-i\phi_{\rm CP}(f_{\rm NE})}
|f_{\rm NE}\rangle.
\end{equation}
The simplest case would be given by semileptonic states of the kind 
$f_{\rm NE}=K^-\ell^+\nu_{\ell}$, where the positive charge of the 
lepton would signal the decay of a $D^0$ meson, as 
$\overline{D^0}\not\to K^-\ell^+\nu_{\ell}$. From an experimental point of
view, these decays are unfortunately affected by large backgrounds due to
semileptonic $B$ decays, which are hard to control. Consequently, we
must rely on final states of the kind $f_{\rm NE}=\pi^+K^-$,
$\rho^+K^-, ...$, where $D^0\to f_{\rm NE}$ and 
$\overline{D^0}\to f_{\rm NE}$ are Cabibbo-favoured and doubly 
Cabibbo-suppressed decay processes, respectively, leading to subtle
interference effects \cite{KaLo,AtSo,ADS}.

\subsection{Amplitudes}
Employing the formalism discussed in Subsection~\ref{subsec:ampl}, we 
obtain
\begin{eqnarray}
\lefteqn{
A(\overline{B^0_q}\to D[\to f_{\rm NE}]f_r)}\nonumber\\
&&=A(\overline{B^0_q}\to D^0f_r)A(D^0\to f_{\rm NE})+
A(\overline{B^0_q}\to\overline{D^0}f_r)
A(\overline{D^0}\to f_{\rm NE})\nonumber\\
&&=\frac{G_{\rm F}}{\sqrt{2}}\overline{v}_r\overline{M}_{f_r}
A(D^0\to f_{\rm NE})\left[1+\frac{v_r}{\overline{v}_r}
a_{f_r}e^{i\delta_{f_r}}r_De^{i\delta_D}\right]
\end{eqnarray}
and
\begin{eqnarray}
\lefteqn{A(B^0_q\to D[\to f_{\rm NE}]f_r)}\nonumber\\
&&=A(B^0_q\to\overline{D^0}f_r)A(\overline{D^0}\to f_{\rm NE})+
A(B^0_q\to D^0f_r)A(D^0\to f_{\rm NE})\nonumber\\
&&=\eta_{f_r}e^{i\phi_{\rm CP}(B_q)}\frac{G_{\rm F}}{\sqrt{2}}
\overline{v}_r^\ast\overline{M}_{f_r}A(D^0\to f_{\rm NE})
\left[r_De^{i\delta_D}+\frac{v_r^\ast}{\overline{v}_r^\ast}
a_{f_r}e^{i\delta_{f_r}}\right],
\end{eqnarray}
where
\begin{equation}
r_De^{i\delta_D}\equiv e^{i\phi_{\rm CP}(D)}
\frac{A(\overline{D^0}\to f_{\rm NE})}{A(D^0\to f_{\rm NE})}.
\end{equation}
Although $r_De^{i\delta_D}$ depends on the considered final state
$f_{\rm NE}$, we do not indicate this through an additional label
for simplicity. Taking now into account (\ref{CKM-fact1}) and 
(\ref{CKM-fact2}), as well as (\ref{X-def}), we eventually arrive at
\begin{equation}\label{xi-NP}
\xi_{D[\to f_{\rm NE}]f_r}^{(q)}\equiv\xi_{f_r}^{(q)}=
-\eta_{f_r}e^{-i\phi_q}\left[\frac{1+e^{-i\gamma}
x_{f_r}e^{i\delta_{f_r}}r_De^{i\delta_D}}{r_De^{i\delta_D}+e^{+i\gamma}
x_{f_r}e^{i\delta_{f_r}}}\right].
\end{equation}

Let us now consider the CP-conjugate state $\overline{f}_{\rm NE}$.
If we assume that CP-violating effects at the $D$-decay amplitude level
are negligibly small, so that the corresponding low-energy effective
Hamiltonian ${\cal H}_{\rm eff}^{(D)}$ satisfies (\ref{CP-D}), we obtain
\begin{displaymath}
A(\overline{D^0}\to f_{\rm NE})=
\langle f_{\rm NE}|{\cal H}_{\rm eff}^{(D)}|\overline{D^0}\rangle
=\langle f_{\rm NE}|({\cal CP})^\dagger({\cal CP})
{\cal H}_{\rm eff}^{(D)}({\cal CP})^\dagger({\cal CP})|\overline{D^0}\rangle
\end{displaymath}
\begin{equation}\label{D-FNE1}
=e^{-i[\phi_{\rm CP}(D)+\phi_{\rm CP}(f_{\rm NE})]}
\langle\overline{f}_{\rm NE}|{\cal H}_{\rm eff}^{(D)\dagger}|D^0\rangle=
e^{-i[\phi_{\rm CP}(D)+\phi_{\rm CP}(f_{\rm NE})]}A(D^0\to 
\overline{f}_{\rm NE}).
\end{equation}
An analogous calculation yields
\begin{equation}\label{D-FNE2}
A(\overline{D^0}\to\overline{f}_{\rm NE})=
e^{-i[\phi_{\rm CP}(D)-\phi_{\rm CP}(f_{\rm NE})]}A(D^0\to f_{\rm NE}).
\end{equation}
Consequently, we obtain
\begin{equation}
r_De^{i\delta_D}=e^{i\phi_{\rm CP}(D)}
\frac{A(\overline{D^0}\to f_{\rm NE})}{A(D^0\to f_{\rm NE})}=
e^{-i\phi_{\rm CP}(D)}\frac{A(D^0\to 
\overline{f}_{\rm NE})}{A(\overline{D^0}\to\overline{f}_{\rm NE})}.
\end{equation}
Taking into account this relation, as well as (\ref{D-FNE2}), the 
$B_q\to D[\to\overline{f}_{\rm NE}]f_r$ decay amplitudes can be written 
as follows:
\begin{eqnarray}
\lefteqn{A(\overline{B^0_q}\to D[\to\overline{f}_{\rm NE}]f_r)}\nonumber\\
&&=A(\overline{B^0_q}\to D^0f_r)A(D^0\to\overline{f}_{\rm NE})+
A(\overline{B^0_q}\to\overline{D^0}f_r)
A(\overline{D^0}\to\overline{f}_{\rm NE})\nonumber\\
&&=\frac{G_{\rm F}}{\sqrt{2}}\overline{v}_r\overline{M}_{f_r}
A(D^0\to f_{\rm NE})e^{i\phi_{\rm CP}(f_{\rm NE})}
\left[r_De^{i\delta_D}+\frac{v_r}{\overline{v}_r}
a_{f_r}e^{i\delta_{f_r}}\right]
\end{eqnarray}
\begin{eqnarray}
\lefteqn{A(B^0_q\to D[\to\overline{f}_{\rm NE}]f_r)}\nonumber\\
&&=A(B^0_q\to\overline{D^0}f_r)A(\overline{D^0}\to\overline{f}_{\rm NE})+
A(B^0_q\to D^0f_r)A(D^0\to\overline{f}_{\rm NE})\nonumber\\
&&=\eta_{f_r}e^{i\phi_{\rm CP}(B_q)}\frac{G_{\rm F}}{\sqrt{2}}
\overline{v}_r^\ast\overline{M}_{f_r}A(D^0\to f_{\rm NE})
e^{i\phi_{\rm CP}(f_{\rm NE})}
\left[1+\frac{v_r^\ast}{\overline{v}_r^\ast}a_{f_r}e^{i\delta_{f_r}}
r_De^{i\delta_D}\right],
\end{eqnarray}
yielding
\begin{equation}\label{xi-NP-bar}
\xi_{D[\to\overline{f}_{\rm NE}]f_r}^{(q)}
\equiv\overline{\xi}_{f_r}^{(q)}=
-\eta_{f_r}e^{-i\phi_q}\left[\frac{r_De^{i\delta_D}+
e^{-i\gamma}x_{f_r}e^{i\delta_{f_r}}}{1+e^{+i\gamma}
x_{f_r}e^{i\delta_{f_r}}r_De^{i\delta_D}}\right].
\end{equation}
Note that the convention-dependent phase $\phi_{\rm CP}(f_{\rm NE})$ 
cancels in this observable. 

It is interesting to note that $\overline{\xi}_{f_r}^{(q)}$ and
$\xi_{f_r}^{(q)}$ satisfy the following relation:
\begin{equation}
\overline{\xi}_{f_r}^{(q)}=\frac{e^{-i2\phi_q}}{\left.\xi_{f_r}^{(q)}
\right|_{\gamma\to-\gamma}}.
\end{equation}
Moreover, we have
\begin{equation}
\left.\overline{\xi}_{f_r}^{(q)}\times\xi_{f_r}^{(q)}\right|_{r_D=0}=
e^{-i2(\phi_q+\gamma)},
\end{equation}
i.e.\ the hadronic parameter $x_{f_r}e^{i\delta_{f_r}}$ would cancel in 
this product in the special case of $r_D=0$, which would correspond to
$f_{\rm NE}=K^-\ell^+\nu_{\ell}$. As we noted above, final states
of this kind are unfortunately extremely challenging from a practical point
of view, so that we have to use $f_{\rm NE}=\pi^+K^-, ...$, where we have 
to care about the $r_D$ effects.

Before we turn to the observables provided by the 
time-dependent $B_q\to D[\to f_{\rm NE}]f_r$ and 
$B_q\to D[\to \overline{f}_{\rm NE}]f_r$ decay rates, let us first 
discuss the $D$-decay parameter $r_De^{i\delta_D}$ in more detail
for the important special case of $f_{\rm NE}=\pi^+K^-$.

\boldmath
\subsection{A Closer Look at $r_De^{i\delta_D}$ for
$f_{\rm NE}=\pi^+K^-$}\label{subsec:rD}
\unboldmath
Let us consider $f_{\rm NE}=\pi^+K^-$ in this subsection to discuss
the corresponding $D$-decay parameter $r_De^{i\delta_D}$ in more detail.
Interestingly, it can be calculated with the help of the $U$-spin flavour 
symmetry of strong interactions \cite{D-wolf,D-Uspin}, which relates 
strange and down quarks to each other in the same manner as ordinary 
isospin relates up and down quarks, i.e.\ through $SU(2)$ transformations.
Following these lines, we obtain
\begin{equation}\label{rD-U-spin}
\left.r_De^{i\delta_D}\right|_{U {\rm \, spin}}=
-\left(\frac{\lambda^2}{1-\lambda^2}\right)=-0.05.
\end{equation}
In order to explore the impact of $U$-spin-breaking corrections to this 
result, we use the factorization approach to deal with the corresponding
decay amplitudes, which gives
\begin{equation}\label{rD-fact}
\left.r_De^{i\delta_D}\right|_{\rm fact}=
-\left(\frac{\lambda^2}{1-\lambda^2}\right)
\left[\frac{M_D^2-M_\pi^2}{M_D^2-M_K^2}\right]
\left[\frac{f_KF_{D\pi}^{(0)}(M_K^2)}{f_\pi F_{DK}^{(0)}(M_\pi^2)}\right]=
-0.06,
\end{equation}
where we have employed the Bauer--Stech--Wirbel (BSW) form factors to 
calculate the numerical value \cite{BSW}. Comparing with (\ref{rD-U-spin}), 
we observe that the factorizable $U$-spin-breaking corrections to this 
expression are at the $20\%$ level, i.e.\ have a moderate impact. 

Fortunately, we may extract $r_D$ from experimental data, since
not only the Cabibbo-favoured mode $D^0\to\pi^+K^-$ has already been 
measured, but also its doubly Cabibbo-suppressed counterpart,
with the following branching ratios \cite{PDG}:
\begin{eqnarray}
\mbox{BR}(D^0\to\pi^-K^+)&=&(1.48\pm0.21)\times10^{-4}\label{D-BR1}\\
\mbox{BR}(D^0\to\pi^+K^-)&=&(3.80\pm0.09)\times10^{-2},\label{D-BR2}
\end{eqnarray}
which yield, if we assume that these modes exhibit no CP violation and
add the errors in quadrature, 
\begin{equation}\label{rD-exp}
r_D=0.062\pm0.005.
\end{equation}
Consequently, the $\overline{D^0},D^0\to \pi^+K^-$ decay parameter $r_D$ 
is already known with an impressive experimental uncertainty of only 
$8\%$. It is remarkable that 
(\ref{rD-exp}) agrees perfectly with $\left.r_D\right|_{\rm fact}=0.06$, 
although the predictions for the corresponding $D$-decay 
branching ratios obtained within the factorization approach differ 
sizeably from the experimental results (see, for example, \cite{D-fact}). 
This observation also gives us confidence in the expectation that 
$a_{f_r}e^{i\delta_{f_r}}$ should not differ dramatically from 
(\ref{a-fact}). Moreover, our considerations also suggest
\begin{equation}\label{delD-exp}
\delta_D\approx180^\circ.
\end{equation}
In particular, it is very plausible to assume that
\begin{equation}\label{sgn-cosDelD}
\cos\delta_D<0,
\end{equation}
which is satisfied for the whole range of $90^\circ<\delta_D<270^\circ$.
For a selection of attempts to estimate $\delta_D$ in a more elaborate way, 
we refer the reader to \cite{D-Uspin,delD-cal}. These theoretical approaches 
are based on various assumptions and hence involve different degrees of 
model dependence. Obviously, it would be very desirable to determine
$\delta_D$ through experiment. To this end, a set of measurements at a 
charm factory were proposed in \cite{charm-factory}, allowing also the
extraction of the $D^0$--$\overline{D^0}$ mixing parameters. In 
Section~\ref{sec:strat}, we will discuss how we may determine 
$\delta_D$ from the $B_q\to Df_r$ observables.

\boldmath
\subsection{Observables of the Time-Dependent Decay Rates}
\unboldmath
In order to calculate the observables that are provided by the
time-dependent rates of the $B_q\to D[\to f_{\rm NE}]f_r$ and 
$B_q\to D[\to\overline{f}_{\rm NE}]f_r$ decay processes, we just have 
to insert (\ref{xi-NP}) and (\ref{xi-NP-bar}) into (\ref{obs-expr}). 
For the coefficients of the $\cos(\Delta M_qt)$ terms arising in 
(\ref{ee6}), we then obtain   
\begin{eqnarray}
\lefteqn{C(B_q\to D[\to f_{\rm NE}]f_r)\equiv C_{f_r}}\nonumber\\
&&=-\frac{1}{N_{f_r}}
\left[(1-r_D^2)(1-x^2_{f_r})+4\,x_{f_r}r_D\sin(\gamma-\delta_D)\sin
\delta_{f_r}\right]\label{C-NP}
\end{eqnarray}
\begin{eqnarray}
\lefteqn{C(B_q\to D[\to\overline{f}_{\rm NE}]f_r)\equiv 
\overline{C}_{f_r}}\nonumber\\
&&=+\frac{1}{\overline{N}_{f_r}}
\left[(1-r_D^2)(1-x^2_{f_r})-4\,x_{f_r}r_D\sin(\gamma+\delta_D)
\sin\delta_{f_r}\right],\label{C-NP-bar}
\end{eqnarray}
with
\begin{equation}\label{Nf-def}
N_{f_r}=(1+r_D^2)(1+x^2_{f_r})+4\,x_{f_r}r_D\cos(\gamma-\delta_D)
\cos\delta_{f_r}
\end{equation}
\begin{equation}\label{Nf-bar-def}
\overline{N}_{f_r}=(1+r_D^2)(1+x^2_{f_r})+4\,x_{f_r}r_D\cos(\gamma+\delta_D)
\cos\delta_{f_r},
\end{equation}
and notice that these observables satisfy the following relation: 
\begin{equation}\label{C-rel}
\overline{C}_{f_r}=-\left.C_{f_r}\right|_{\gamma\to-\gamma}. 
\end{equation}
In order to deal with the interference effects between the Cabibbo-favoured
and doubly Cabibbo-suppressed $D$ decays, which are described by $r_D$, 
we expand (\ref{C-NP}) and (\ref{C-NP-bar}) in this parameter and
keep only the leading ${\cal O}(r_D)$ corrections. On the other hand, we 
keep {\it all} orders in $x_{f_r}$. Following these lines, we arrive at
\begin{equation}\label{C-expand}
C_{f_r}=-\left[\frac{1-x^2_{f_r}}{1+x^2_{f_r}}\right]+
\frac{4\,x_{f_r}}{(1+x^2_{f_r})^2}\left[\cos(\gamma-\delta_D+\delta_{f_r})-
x^2_{f_r}\cos(\gamma-\delta_D-\delta_{f_r})\right]r_D + {\cal O}(r_D^2).
\end{equation}
The corresponding expression for $\overline{C}_{f_r}$ can be obtained 
straightforwardly from (\ref{C-expand}) with the help of (\ref{C-rel}). 
For the following considerations, it is convenient to introduce
\begin{equation}\label{Ctilde-p}
\langle \tilde C_{f_r} \rangle_+\equiv\frac{\overline{C}_{f_r}+C_{f_r}}{2}=
\frac{4\,x_{f_r}\sin\gamma}{(1+x^2_{f_r})^2}\left[\sin(\delta_D-\delta_{f_r})
-x^2_{f_r}\sin(\delta_D+\delta_{f_r})\right]r_D + {\cal O}(r_D^2)
\end{equation}
\begin{eqnarray}
\lefteqn{\langle \tilde C_{f_r} \rangle_-\equiv
\frac{\overline{C}_{f_r}-C_{f_r}}{2}}\nonumber\\
&&=\left[\frac{1-x^2_{f_r}}{1+x^2_{f_r}}\right]-
\frac{4\,x_{f_r}\cos\gamma}{(1+x^2_{f_r})^2}\left[\cos(\delta_D-\delta_{f_r})
-x^2_{f_r}\cos(\delta_D+\delta_{f_r})\right]r_D + 
{\cal O}(r_D^2).\label{Ctilde-m}
\end{eqnarray}
Note that $\langle \tilde C_{f_r} \rangle_+$ arises at the $r_D$ level,
i.e.\ $\langle \tilde C_{f_r} \rangle_+={\cal O}(r_D)$, whereas
$\langle \tilde C_{f_r} \rangle_-={\cal O}(1)$.

As far as the coefficients of the $\sin(\Delta M_qt)$ terms in (\ref{ee6})
are concerned, we have
\begin{eqnarray}\label{S-NP}
\lefteqn{S(B_q\to D[\to f_{\rm NE}]f_r)\equiv S_{f_r}=
\frac{2\,\eta_{f_r}}{N_{f_r}}
\Bigl[x_{f_r}\sin(\phi_q+\gamma+\delta_{f_r})}\nonumber\\
&&+r_D\left\{\sin(\phi_q+\delta_D)+x_{f_r}^2\sin(\phi_q+2\gamma-\delta_D)+
x_{f_r}r_D\sin(\phi_q+\gamma-\delta_{f_r})\right\}\Bigr]
\end{eqnarray}
\begin{eqnarray}\label{S-NP-bar}
\lefteqn{S(B_q\to D[\to\overline{f}_{\rm NE}]f_r)\equiv \overline{S}_{f_r}=
\frac{2\,\eta_{f_r}}{\overline{N}_{f_r}}
\Bigl[x_{f_r}\sin(\phi_q+\gamma-\delta_{f_r})}\nonumber\\
&&+r_D\left\{\sin(\phi_q-\delta_D)+x_{f_r}^2\sin(\phi_q+2\gamma+\delta_D)+
x_{f_r}r_D\sin(\phi_q+\gamma+\delta_{f_r})\right\}\Bigr],
\end{eqnarray}
where $N_{f_r}$ and $\overline{N}_{f_r}$ were introduced in 
(\ref{Nf-def}) and (\ref{Nf-bar-def}), respectively. 
We notice that these observables are related to each other through
\begin{equation}\label{S-rel}
\overline{S}_{f_r}=\left.S_{f_r}\right|_{\delta_{f_r}
\to-\delta_{f_r}, \delta_D\to-\delta_D}.
\end{equation}
Keeping again only the leading-order terms in $r_D$, but taking into account 
{\it all} orders in $x_{f_r}$, we obtain
\begin{eqnarray}\label{S-NP-exp}
S_{f_r}&=&\frac{2\,\eta_{f_r}}{1+x^2_{f_r}}\Biggl[x_{f_r}
\sin(\phi_q+\gamma+\delta_{f_r})+r_D\biggl\{\sin(\phi_q+\delta_D)
+x_{f_r}^2\sin(\phi_q+2\gamma-\delta_D)\nonumber\\
&&-\frac{4\,x_{f_r}^2}{1+x^2_{f_r}}
\cos\delta_{f_r}\cos(\gamma-\delta_D)\sin(\phi_q+\gamma+\delta_{f_r})
\biggr\}\Biggr]+{\cal O}(r_D^2).
\end{eqnarray}
The corresponding expression for $\overline{S}_{f_r}$ can be obtained 
straightforwardly from (\ref{S-NP-exp}) with the help of (\ref{S-rel}). 
In analogy to (\ref{Ctilde-p}) and (\ref{Ctilde-m}), we introduce
\begin{eqnarray}
\lefteqn{\langle \tilde S_{f_r} \rangle_+\equiv
\frac{\overline{S}_{f_r}+S_{f_r}}{2}=\frac{2\,\eta_{f_r}}{1+x^2_{f_r}}\Biggl[
x_{f_r}\sin(\phi_q+\gamma)\cos\delta_{f_r}+r_D\Biggl\{\cos\delta_D
\biggl(\sin\phi_q}\nonumber\\
&&+x^2_{f_r}\sin(\phi_q+2\gamma)\biggr)-\frac{4\,x_{f_r}^2}{1+x^2_{f_r}}\cos
\delta_{f_r}\biggl(\cos\delta_D\cos\delta_{f_r}\cos\gamma\sin(\phi_q+\gamma)
\nonumber\\
&&+\sin\delta_D\sin\delta_{f_r}\sin\gamma\cos(\phi_q+\gamma)\biggr)\Biggr\}
\Biggr] +{\cal O}(r_D^2)\label{tilS+exp}
\end{eqnarray}
and
\begin{eqnarray}
\lefteqn{\langle \tilde S_{f_r} \rangle_-\equiv
\frac{\overline{S}_{f_r}-S_{f_r}}{2}=-\frac{2\,\eta_{f_r}}{1+x^2_{f_r}}\Biggl[
x_{f_r}\cos(\phi_q+\gamma)\sin\delta_{f_r}+r_D\Biggl\{\sin\delta_D
\biggl(\cos\phi_q}\nonumber\\
&&-x^2_{f_r}\cos(\phi_q+2\gamma)\biggr)-\frac{4\,x_{f_r}^2}{1+x^2_{f_r}}\cos
\delta_{f_r}\biggl(\cos\delta_D\sin\delta_{f_r}\cos\gamma\cos(\phi_q+\gamma)
\nonumber\\
&&+\sin\delta_D\cos\delta_{f_r}\sin\gamma\sin(\phi_q+\gamma)\biggr)\Biggr\}
\Biggr] +{\cal O}(r_D^2),\label{tilS-exp}
\end{eqnarray}
respectively. For an efficient determination of $\gamma$ from these 
observables, the untagged rate asymmetry (\ref{Gpm-untagged}) of the $B_q\to
D_\pm f_r$ case, providing the quantity $\Gamma_{+-}^{f_r}$, is again an 
important ingredient. Interestingly, the decays of the kind 
$B_q\to D[\to f_{\rm NE}] f_r$ offer another useful untagged rate 
asymmetry, which is the subject of the following subsection.

\boldmath
\subsection{Untagged Observables}
\unboldmath
If we measure untagged decay rates of the kind specified in 
(\ref{untagged}), we may extract the following unevolved, 
i.e.\ time-independent, untagged rates:
\begin{eqnarray}
\langle\Gamma(B_q\to D[\to f_{\rm NE}]f_r)\rangle&=&
\Gamma(B_q^0\to D[\to f_{\rm NE}]f_r)+
\Gamma(\overline{B_q^0}\to D[\to f_{\rm NE}]f_r)\nonumber\\
&=&\Gamma(\overline{B^0_q}\to D^0f_r)\Gamma(D^0\to f_{\rm NE})N_{f_r}
\end{eqnarray}
\begin{eqnarray}
\langle\Gamma(B_q\to D[\to \overline{f}_{\rm NE}]f_r)\rangle&=&
\Gamma(B_q^0\to D[\to \overline{f}_{\rm NE}]f_r)+
\Gamma(\overline{B_q^0}\to D[\to \overline{f}_{\rm NE}]f_r)\nonumber\\
&=&\Gamma(\overline{B^0_q}\to D^0f_r)\Gamma(D^0\to f_{\rm NE})
\overline{N}_{f_r},
\end{eqnarray}
where $N_{f_r}$ and $\overline{N}_{f_r}$ were introduced in 
(\ref{Nf-def}) and (\ref{Nf-bar-def}), respectively. Consequently,
we obtain
\begin{eqnarray}
\tilde\Gamma_{f_r}&\equiv&
\frac{\langle\Gamma(B_q\to D[\to \overline{f}_{\rm NE}]f_r)\rangle-
\langle\Gamma(B_q\to D[\to f_{\rm NE}]f_r)\rangle}{\langle
\Gamma(B_q\to D[\to \overline{f}_{\rm NE}]f_r)\rangle+
\langle\Gamma(B_q\to D[\to f_{\rm NE}]f_r)\rangle}\nonumber\\
&=&-\left[\frac{4\,x_{f_r}\cos\delta_{f_r} \sin\gamma\,r_D 
\sin\delta_D}{(1+r_D^2)(1+x^2_{f_r})+4\,x_{f_r}\cos\delta_{f_r}\cos\gamma\,
r_D\cos\delta_D}\right].\label{Gam-NE-1}
\end{eqnarray}
Note that $\Gamma(\overline{B^0_q}\to D^0f_r)$ and
$\Gamma(D^0\to f_{\rm NE})$ cancel in this untagged rate asymmetry.
Using (\ref{Gpm-untagged}), we may eliminate $x_{f_r}$ and $\delta_{f_r}$,
and arrive at the following {\it exact} relation:
\begin{equation}\label{G-NE-exact}
\tilde\Gamma_{f_r}=-\left[\frac{2\,\Gamma_{+-}^{f_r}r_D\sin\delta_D}{1+
2\,\Gamma_{+-}^{f_r}r_D\cos\delta_D+r_D^2}\right]\tan\gamma,
\end{equation}
which allows us to probe $r_D\sin\delta_D$ nicely through
\begin{equation}\label{sinDelD}
\frac{\tilde\Gamma_{f_r}}{\Gamma_{+-}^{f_r}}=
-2\,r_D\sin\delta_D\tan\gamma + {\cal O}(r_D^2),
\end{equation}
i.e.\ with the help of untagged $B_q\to Df_r$ rate asymmetries. Following
these lines, we may in particular check whether (\ref{delD-exp}) is 
actually satisfied. We shall come back to this important point below.

\boldmath
\section{Efficient Extraction of $\gamma$ and Hadronic 
Parameters}\label{sec:strat}
\setcounter{equation}{0}
\unboldmath
Let us now focus on the extraction of the angle $\gamma$ of the unitarity
triangle, as well as interesting hadronic parameters. Before turning to
the general case, it is very instructive to consider first $r_D=0$.

\boldmath
\subsection{Case of $r_D=0$}\label{sec:LO}
\unboldmath
If we neglect the $r_D$ terms, expression (\ref{tilS+exp}) for 
$\langle \tilde S_{f_r} \rangle_+$ simplifies considerably:
\begin{equation}\label{LO-Sp}
\left.\eta_{f_r}\langle \tilde S_{f_r} \rangle_+\right|_{r_D=0}
=\left[\frac{2\,x_{f_r}
\cos\delta_{f_r}}{1+x^2_{f_r}}\right]\sin(\phi_q+\gamma).
\end{equation}
If we compare with (\ref{Gpm-untagged}), we observe that the hadronic factor
on the right-hand side of this expression can straightforwardly be eliminated 
with the help of $\Gamma_{+-}^{f_r}$, yielding
\begin{equation}\label{S-LO}
\left.\eta_{f_r}\langle \tilde S_{f_r} \rangle_+\right|_{r_D=0}=
\Gamma_{+-}^{f_r}\left[\sin\phi_q+\tan\gamma\cos\phi_q\right].
\end{equation}
Consequently, we obtain the following simple relation: 
\begin{equation}\label{tan-gam-LO}
\tan\gamma\cos\phi_q=
\left[\frac{\eta_{f_r}\langle \tilde S_{f_r} \rangle_+}{\Gamma_{+-}^{f_r}}
\right]-\sin\phi_q + {\cal O}(r_D),
\end{equation}
which is the counterpart of (\ref{tan-gam-CP}), and allows an efficient
and essentially unambiguous determination of $\gamma$. It should be 
emphasized again that (\ref{tan-gam-CP}) is an {\it exact} relation, 
whereas (\ref{tan-gam-LO}) is affected by corrections of ${\cal O}(r_D)$, 
which are discussed in detail in Subsection~\ref{sec:NLO}.

In order to determine the strong phase $\delta_{f_r}$, we may use
\begin{equation}\label{LO-Sm}
\left.\eta_{f_r}\langle \tilde S_{f_r} \rangle_-\right|_{r_D=0}
=-\left[\frac{2\,x_{f_r}\sin\delta_{f_r}}{1+x^2_{f_r}}\right]
\cos(\phi_q+\gamma),
\end{equation}
where $x_{f_r}$ can again be eliminated  through $\Gamma_{+-}^{f_r}$,
yielding
\begin{equation}
\left.\eta_{f_r}\langle \tilde S_{f_r} \rangle_-\right|_{r_D=0}
=-\Gamma_{+-}^{f_r}\left[\cos\phi_q-
\tan\gamma\sin\phi_q\right]\tan\delta_{f_r}.
\end{equation}
Employing (\ref{tan-gam-LO}) to fix $\tan\gamma$, we arrive at
\begin{equation}\label{tan-del-det-LO}
\tan\delta_{f_r}=-\left[\frac{\eta_{f_r}\langle \tilde S_{f_r} \rangle_-
\cos\phi_q}{\Gamma_{+-}^{f_r}-\eta_{f_r}\langle \tilde S_{f_r} \rangle_+
\sin\phi_q}\right]+{\cal O}(r_D).
\end{equation}
As we have seen in Subsection~\ref{subsec:ampl}, it is plausible to
assume that $\cos\delta_{f_r}<0$, which allows us to fix $\delta_{f_r}$ 
{\it unambiguously} through (\ref{tan-del-det-LO}). In 
particular, we may distinguish between $\delta_{f_r}>180^\circ$ and 
$\delta_{f_r}<180^\circ$. Finally, (\ref{Ctilde-m}) implies
\begin{equation}\label{LO-Cm}
\left.\langle \tilde C_{f_r} \rangle_-\right|_{r_D=0}=
\left[\frac{1-x_{f_r}^2}{1+x_{f_r}^2}\right],
\end{equation}
which allows us to extract $x_{f_r}$ through 
\begin{equation}\label{x-det-LO}
|x_{f_r}|=\sqrt{\frac{1-\langle \tilde C_{f_r} \rangle_-}{1+
\langle \tilde C_{f_r} \rangle_-}} + {\cal O}(r_D),
\end{equation}
where $x_{f_r}$ is positive (negative) for $r=s$ ($d$), as can be 
seen in (\ref{X-def}).

The expressions given in this subsection are valid to {\it all} orders
in $x_{f_r}$, but neglect terms of ${\cal O}(r_D)$. Let us discuss next
how these corrections can be taken into account in the extraction of
$\gamma$ and the hadronic parameters $\delta_{f_r}$ and $x_{f_r}$. Because 
of the experimental value given in (\ref{rD-exp}), we expect the $r_D$ 
effects to play a minor r\^ole, provided the observables on the 
left-hand sides of (\ref{LO-Sp}), (\ref{LO-Sm}) and (\ref{LO-Cm}) 
are found to be significantly 
larger than ${\cal O}(0.1)$. For the numerical examples discussed in
Subsection~\ref{sec:examp}, using plausible input parameters, this is 
actually the case for $\langle \tilde S_{f_r} \rangle_+$ and 
$\langle \tilde C_{f_r} \rangle_-$, so that (\ref{tan-gam-LO}) and 
(\ref{x-det-LO}) are found to be good approximations
for the extraction of $\gamma$ and $x_{f_r}$, respectively. On the other 
hand, $\langle \tilde S_{f_r} \rangle_-$ takes small values at the 0.1 
level, so that (\ref{tan-del-det-LO}) would only allow a poor 
determination of $\delta_{f_r}$ in these examples. For a precision analysis, 
of course, it is anyway crucial to take the $r_D$ effects into account, 
also in the extractions of $\gamma$ and $x_{f_r}$.

\boldmath
\subsection{Inclusion of the $r_D$ Effects}\label{sec:NLO}
\unboldmath
As we have seen above, $\langle \tilde S_{f_r} \rangle_+$ is the key
observable for an efficient determination of $\gamma$. In order to include
the $r_D$ effects, we use (\ref{tilS+exp}) as a starting point. If we 
employ now (\ref{Gpm-untagged}), (\ref{Ctilde-m}), (\ref{tilS-exp}) and 
(\ref{Gam-NE-1}) to deal with the hadronic decay parameters and neglect 
again the ${\cal O}(r_D^2)$ terms, but take into account {\it all} orders 
in $x_{f_r}$, we eventually arrive at the following generalization of 
(\ref{S-LO}):
\begin{eqnarray}
\eta_{f_r}\langle \tilde S_{f_r} \rangle_+&=&
\Gamma_{+-}^{f_r}\Bigl[\sin\phi_q+\tan\gamma\cos\phi_q\Bigr]
\Big[1-2\,\Gamma_{+-}^{f_r}r_D\cos\delta_D\Bigr]-
\tilde\Gamma_{f_r}\,\eta_{f_r}\langle \tilde S_{f_r}\rangle_-\nonumber\\
&&+2\,r_D\cos\delta_D\left[\sin\phi_q+(1-\langle \tilde C_{f_r} \rangle_-)
\cos(\phi_q+\gamma)\sin\gamma\right]+{\cal O}(r_D^2).\label{S-NLO}
\end{eqnarray}
Note that $\tilde\Gamma_{f_r}={\cal O}(r_D)$, and that (\ref{S-NLO}) 
is an {\it exact} relation. Interestingly, as in the case of 
(\ref{S-LO}), we can still eliminate the hadronic parameters $x_{f_r}$ 
and $\delta_{f_r}$ in an elegant manner, although the resulting 
expression is now more complicated, involving -- in addition to 
$\Gamma_{+-}^{f_r}$ -- also the observables $\tilde\Gamma_{f_r}$, 
$\langle \tilde S_{f_r}\rangle_-$ and $\langle \tilde C_{f_r} \rangle_-$.

Let us assume, in the following, that $r_D$ has been determined through
$D$-decay studies. In particular, we shall have the important case of 
$f_{\rm NE}=\pi^+K^-$ in mind, where (\ref{D-BR1}) and (\ref{D-BR2}) yield 
the result given in (\ref{rD-exp}), which has already an impressive accuracy. 
By the time the observables of the $B$ decays considered in this paper can 
be measured, the experimental accuracy of $r_D$ will 
have increased a lot. As far as the strong phase 
$\delta_D$ is concerned, we have seen in Subsection~\ref{subsec:rD} that 
it is plausible to assume $\delta_D\approx180^\circ$. Interestingly, 
$r_D$ enters in (\ref{S-NLO}) only in combination with $\cos\delta_D$. 
Consequently, we could assume $\cos\delta_D\approx-1$ as a first 
approximation, which is rather stable under deviations of $\delta_D$ from 
$180^\circ$. However, we may actually do much better with the help of 
(\ref{sinDelD}). Since $\tan\gamma$ is multiplied here by $r_D$, we 
can simply fix it through the leading-order expression
(\ref{tan-gam-LO}), yielding
\begin{equation}\label{rD-sinD}
r_D\sin\delta_D=-\left[\frac{\tilde\Gamma_{f_r}\cos\phi_q}{2\left(
\eta_{f_r}\langle \tilde S_{f_r} \rangle_+-\Gamma_{+-}^{f_r}\sin\phi_q
\right)}\right]+{\cal O}(r_D^2),
\end{equation}
which implies
\begin{equation}\label{rD-cosD}
r_D\cos\delta_D=-\sqrt{r_D^2-\left[\frac{\tilde\Gamma_{f_r}
\cos\phi_q}{2\left(\eta_{f_r}\langle \tilde S_{f_r} \rangle_+-
\Gamma_{+-}^{f_r}\sin\phi_q\right)}\right]^2}+{\cal O}(r_D^2),
\end{equation}
where we have chosen the sign in front of the square root to be in 
accordance with (\ref{sgn-cosDelD}). Consequently, if 
$\tilde\Gamma_{f_r}$ should be found experimentally
to be negligibly small, (\ref{rD-cosD}) gives $r_D\cos\delta_D=-1$,
as na\"\i vely expected. On the other hand, if a sizeable value of
$\tilde\Gamma_{f_r}$ should be measured, (\ref{rD-sinD}) and 
(\ref{rD-cosD}) allow us to determine $\delta_D$ {\it unambiguously},
thereby distinguishing in particular whether this phase is smaller or 
larger than $180^\circ$; we will illustrate this determination in 
Subsection~\ref{sec:examp}. Finally, we may insert (\ref{rD-cosD})
into our central expression (\ref{S-NLO}), allowing us to then take 
into account the ${\cal O}(r_D)$ effects completely in the extraction 
of $\gamma$. 

Before discussing this in more detail in the next subsection, let us 
first give other useful expressions: 
\begin{eqnarray}
\eta_{f_r}\langle \tilde S_{f_r} \rangle_-&=&
-\Gamma_{+-}^{f_r}\Bigl[\cos\phi_q-\tan\gamma\sin\phi_q\Bigr]
\Big[1-2\,\Gamma_{+-}^{f_r}r_D\cos\delta_D\Bigr]\tan\delta_{f_r}
-\tilde\Gamma_{f_r}\,\eta_{f_r}\langle \tilde S_{f_r}\rangle_+\nonumber\\
&&-2\,r_D\sin\delta_D\left[\cos\phi_q-(1-\langle \tilde C_{f_r} \rangle_-)
\cos(\phi_q+\gamma)\cos\gamma\right]+{\cal O}(r_D^2)\label{SmNLO}\\
\langle \tilde C_{f_r} \rangle_-&=&\left[\frac{1-x^2_{f_r}}{1+x^2_{f_r}}
\right]\Big[1-2\,\Gamma_{+-}^{f_r}r_D\cos\delta_D\Bigr]-
2\,r_D\sin\delta_D\Big[\Gamma_{+-}^{f_r}\tan\delta_{f_r}\Bigr]+
{\cal O}(r_D^2),\label{CmNLO}
\end{eqnarray}
which involve -- in contrast to (\ref{S-NLO}) -- the hadronic
parameters $\delta_{f_r}$ and $x_{f_r}$. Consequently, 
$\langle \tilde S_{f_r} \rangle_-$ and $\langle \tilde C_{f_r} \rangle_-$
allow us to determine these interesting quantities. Finally, we have
\begin{equation}\label{CpNLO}
\langle \tilde C_{f_r} \rangle_+=-\tilde\Gamma_{f_r}
\langle \tilde C_{f_r} \rangle_-
-2\,r_D\cos\delta_D\Big[\Gamma_{+-}^{f_r}\tan\delta_{f_r}\tan\gamma\Bigr]+
{\cal O}(r_D^2).
\end{equation}

\boldmath
\subsection{Main Strategy and Resolution of Discrete 
Ambiguities}\label{subsec:main-strat}
\unboldmath
Since interference effects between the $\overline{B^0_q}\to
\overline{D^0}f_s$ and $\overline{B^0_q}\to D^0f_s$ decay paths are large, 
in contrast to the case of $\overline{B^0_q}\to \overline{D^0}f_d$ and 
$\overline{B^0_q}\to D^0f_d$, the former channels appear much more 
attractive for the extraction of $\gamma$, despite the smaller 
$\overline{B^0_q}\to D^0 f_s$ branching ratios, which are suppressed
relative to the $\overline{B^0_q}\to D^0 f_d$ case by $\lambda^2$. 
Let us therefore focus now on decays of the kind $B_d\to DK_{\rm S(L)}$ and 
$B_s\to D\eta^{(')}, D\phi, ...$, i.e.\ on the $r=s$ case. We shall 
briefly come back to the $r=d$ modes in Subsection~\ref{sec:r-d-case}.
Concerning an efficient and essentially unambiguous extraction of $\gamma$, 
we may proceed along the following steps:
\begin{itemize}
\item The most straightforward measurement is the determination of 
$\Gamma_{+-}^{f_s}$ from untagged $B_q\to D_\pm f_s$ rates with the help of
(\ref{Gpm-untagged}). As pointed out in \cite{RF-BD-CP}, already
this observable provides useful information about $\gamma$, allowing us
to constrain this angle through
\begin{equation}\label{gam-bound}
|\cos\gamma|\geq |\Gamma_{+-}^{f_s}|.
\end{equation}
Moreover, making use of the plausible assumption (\ref{dyn-assumpt}), 
we may fix the sign of $\cos\gamma$ through
\begin{equation}\label{sgn-cos-gam}
\mbox{sgn}(\cos\gamma)=-\mbox{sgn}(\Gamma_{+-}^{f_s}).
\end{equation}
At this point, it is instructive to have a brief look at the situation 
in the $\overline{\rho}$--$\overline{\eta}$ plane illustrated in
Fig.~\ref{fig:UT}. Because of the simple relation
\begin{equation}
\overline{\rho}=R_b\cos\gamma,
\end{equation}
$\mbox{sgn}(\Gamma_{+-}^{f_s})$ allows us to answer the question of whether 
the Wolfenstein parameter $\overline{\rho}$ is positive or negative. 
This is a particularly interesting issue, as the lower bounds to date
on the $B_s$ mass difference $\Delta M_s$ favour 
$\overline{\rho}>0$ \cite{RF-PHYS-REP}.

For the {\it determination} of $\gamma$, we may complement $\Gamma_{+-}^{f_s}$ 
either with a measurement of $\langle S_{f_s}\rangle_\pm$, as discussed in 
detail in \cite{RF-BD-CP} (see also Subsection~\ref{subsec:appl-CP}), or 
-- which is the topic of this section -- through the 
$B_q\to D[\to f_{\rm NE}] f_s$ observables introduced above.

\item Once the mixing-induced observables $\overline{S}_{f_s}$ and 
$S_{f_s}$ have been measured, we may calculate their average 
$\langle\tilde S_{f_s}\rangle_+$, and may determine $\tan\gamma\cos\phi_q$ 
in a very efficient way from (\ref{tan-gam-LO}), neglecting the 
${\cal O}(r_D)$ corrections as a first approximation. It is plausible 
to assume that $\phi_q$ will be known unambiguously through measurements of
$\sin\phi_q$ and $\mbox{sgn}(\cos\phi_q)$ by the time 
$\langle \tilde S_{f_s}\rangle_+$ is accessible. We may then determine 
$\tan\gamma$ unambiguously with the help of (\ref{tan-gam-LO}), 
implying $\gamma=\gamma_1\lor\gamma_2$, where we may choose
$\gamma_1\in[0^\circ,180^\circ]$ and $\gamma_2=\gamma_1+180^\circ$. 
Since $\cos\gamma_1$ and $\cos\gamma_2$ have opposite signs, 
(\ref{sgn-cos-gam}) allows us to distinguish between these two solutions 
for $\gamma$, thereby fixing this angle {\it unambiguously}. 

In this context, it is important to note that the Standard-Model 
interpretation of the observable $\varepsilon_K$, which measures
the ``indirect'' CP violation in the neutral kaon system, implies 
$\gamma\in[0^\circ,180^\circ]$, if we make the very 
plausible assumption that a certain ``bag'' parameter $B_K$ is positive. 
Let us note that we have also assumed implicitly in (\ref{phi-q-def}) 
that another ``bag'' parameter $B_{B_q}$, which is the $B_q$-meson counterpart 
of $B_K$, is positive as well \cite{RF-PHYS-REP}. Indeed, all existing 
non-perturbative methods give positive values for these parameters (for a 
discussion of the very unlikely $B_K<0$, $B_{B_q}<0$ cases, see \cite{GKN}). 
If we follow these lines and assume that $\gamma$ lies between $0^\circ$ 
and $180^\circ$, the knowledge of $\sin\phi_q$ is sufficient for 
an unambiguous extraction of $\gamma$. In this case, which corresponds to
$\overline{\eta}>0$ because of
\begin{equation}
\overline{\eta}=R_b\sin\gamma, 
\end{equation}
we have $\mbox{sgn}(\tan\gamma)=\mbox{sgn}(\cos\gamma)$, and may hence fix 
the sign of $\tan\gamma$ through (\ref{sgn-cos-gam}). Consequently, if
we determine the sign of the right-hand side of (\ref{tan-gam-LO}),
we may also fix the sign of $\cos\phi_q$, which allows us to resolve
the twofold ambiguity arising in the extraction of $\phi_q$ from
$\sin\phi_q$. Finally, we may determine $\gamma$ unambiguously with 
the help of (\ref{tan-gam-LO}). However, since the expectation 
$\gamma\in[0^\circ,180^\circ]$ relies on the Standard-Model interpretation 
of $\varepsilon_K$, it may no longer be correct in the presence of new 
physics. It is hence extremely interesting to decide whether $\gamma$ 
is actually smaller than $180^\circ$, following the strategy discussed 
above or the one proposed in \cite{RF-BD-CP}.

If we apply (\ref{tan-del-det-LO}), we may also extract $\delta_{f_s}$ 
(see, however, the remark at the end of Subsection~\ref{sec:LO}). Once 
a measurement of the observables $\overline{C}_{f_s}$, $C_{f_s}$ 
associated with the $\cos(\Delta M_qt)$ terms is available, which is a bit 
harder than the one of $\overline{S}_{f_s}$, $S_{f_s}$ associated with 
the $\sin(\Delta M_qt)$ terms, we may determine 
$x_{f_s}$ with the help of (\ref{x-det-LO}).

\item Once the experimental accuracy of the mixing-induced observables
$\overline{S}_{f_s}$, $S_{f_s}$ reaches the $2\,r_D\cos\delta_D\approx0.1$ 
level, we have to care about the $r_D$ effects, which can be
taken into account with the help of (\ref{S-NLO}), where $r_D\cos\delta_D$
is fixed through (\ref{rD-cosD}). To this end, $r_D$ has to be determined 
through $D$-decay studies, and $\tilde\Gamma_{f_r}$ has to be extracted 
from untagged $B_q\to D[\to f_{\rm NE}]f_s$ measurements. In contrast to 
(\ref{tan-gam-LO}), we cannot solve (\ref{S-NLO}) straightforwardly for 
$\tan\gamma$. However, we may still determine $\gamma$ in an elegant manner, 
if we use the right-hand side of (\ref{S-NLO}) to calculate 
$\eta_{f_r}\langle \tilde S_{f_r} \rangle_+$ as a function
of this angle. The measured value of $\langle \tilde S_{f_r} \rangle_+$
fixes then $\gamma$, and (\ref{sgn-cos-gam}) allows us again
to resolve the twofold ambiguity. 

Using then (\ref{SmNLO}) with (\ref{rD-sinD}) and (\ref{rD-cosD}), 
we may extract $\tan\delta_{f_s}$, which yields a twofold solution for 
$\delta_{f_s}$; using (\ref{dyn-assumpt}), this ambiguity can be resolved, 
thereby fixing $\delta_{f_s}$ unambiguously. Finally, we may employ
(\ref{CmNLO}) to determine $x_{f_s}$. Needless to note, $\delta_{f_s}$ 
and $x_{f_s}$ provide valuable insights into hadron dynamics. In 
particular, we may check whether the plausible
expectations $\delta_{f_s}\approx180^\circ$ and $x_{f_s}\approx R_b\approx0.4$
are actually satisfied. The tiny observable $\langle \tilde C_{f_r} \rangle_+$,
which originates at the $r_D$ level and is not needed in our strategy, 
offers a consistency check through (\ref{CpNLO}).

\item An interesting playground is also offered if we combine $B_d$- and 
$B_s$-meson decays, which can be done in various ways. For instance, we may 
use the $B_d\to D K_{\rm S(L)}$ 
observables to determine $r_D\cos\delta_D$ from (\ref{S-NLO}) as a function 
of $\gamma$. Using once again (\ref{S-NLO}), but applying it now
to a $B_s$ channel, for example to $B_s\to D \phi$, we may calculate the
corresponding observable $\langle\tilde S_{f_s}\rangle_+$ as a function of 
$\gamma$. The measured $B_s\to D \phi$ observable 
$\langle\tilde S_{f_s}\rangle_+$ then allows us to determine both $\gamma$
and $r_D\cos\delta_D$. Using (\ref{sinDelD}), we may also extract 
$r_D\sin\delta_D$, allowing us to fix $r_D$ and $\delta_D$. Consequently, no 
experimental $D$-meson input is required if we follow this avenue. Finally,
applying (\ref{SmNLO}) and (\ref{CmNLO}) to the $B_d\to D K_{\rm S(L)}$,
$B_s\to D \phi$ channels, we may determine their decay parameters 
$\delta_{f_s}$ and $x_{f_s}$, as discussed above. 

The $B_s$-meson decays are not accessible at the $e^+e^-$ $B$ factories
operating at the $\Upsilon(4S)$ resonance, i.e.\ at BaBar and Belle. However, 
they can be explored at hadronic $B$-decay experiments, in particular 
at LHCb and BTeV. Detailed feasibility studies are strongly encouraged to 
find out which channels are best suited from an experimental point of view 
to implement the strategies discussed above. 
\end{itemize}

\boldmath
\subsection{Special Cases Related to $\Gamma_{+-}^{f_s}=0$}\label{subsec:spec}
\unboldmath
Let us, before we turn to numerical examples, discuss two special cases, 
where $\Gamma_{+-}^{f_s}$ vanishes. The first one corresponds to 
$\gamma=90^\circ \lor 270^\circ$, yielding
\begin{eqnarray}
\eta_{f_s}\langle \tilde S_{f_s} \rangle_+&=&
\left[\frac{2\,x_{f_s}\cos\delta_{f_s}\cos\phi_q}{1+x_{f_s}^2}\right]
\mbox{sgn}(\sin\gamma)-\tilde\Gamma_{f_s}\,\eta_{f_s}\langle \tilde 
S_{f_s}\rangle_-\nonumber\\
&&+2\,r_D\cos\delta_D\sin\phi_q\langle \tilde C_{f_s}\rangle_-
+{\cal O}(r_D^2).\label{Sp-special}
\end{eqnarray}
If we insert $\gamma=90^\circ \lor 270^\circ$ into (\ref{Gam-NE-1})
and use (\ref{Sp-special}) to eliminate the hadronic factor involving
$x_{f_s}$ and $\delta_{f_s}$, we obtain
\begin{equation}
r_D\sin\delta_D=-\left[\frac{\tilde\Gamma_{f_s}
\cos\phi_q}{2\,\eta_{f_s}\langle \tilde S_{f_s} \rangle_+}
\right]+{\cal O}(r_D^2),
\end{equation}
which implies that $r_D\cos\delta_D$ can be fixed with the help of
\begin{equation}\label{cosD-det2}
r_D\cos\delta_D=-\sqrt{r_D^2-\left[\frac{\tilde\Gamma_{f_s}
\cos\phi_q}{2\,\eta_{f_s}\langle \tilde S_{f_s} \rangle_+}
\right]^2}+{\cal O}(r_D^2).
\end{equation}
Here we have determined the sign in front of the square root through
(\ref{sgn-cosDelD}), as in (\ref{rD-cosD}). Consequently, should we
measure a vanishing value of $\Gamma_{+-}^{f_s}$, and should we
simultaneously find a non-vanishing value for the term in square brackets in 
(\ref{Sp-special}), thereby indicating $\cos\delta_{f_s}\not=0$, we 
would conclude that $\gamma=90^\circ \lor 270^\circ$. Since the 
${\cal O}(r_D)$ corrections generally play a minor r\^ole, a measurement 
of a large non-vanishing value of $\langle \tilde S_{f_s} \rangle_+$ would
be sufficient to this end. If we then use (\ref{dyn-assumpt}) and assume 
that we know the sign of $\cos\phi_q$, (\ref{Sp-special}) allows us to 
fix the sign of $\sin\gamma$, thereby distinguishing between 
$\gamma=90^\circ$ and $270^\circ$. In the case of 
$\gamma=90^\circ \lor 270^\circ$, the ${\cal O}(r_D)$ corrections to 
(\ref{CmNLO}) vanish, i.e.
\begin{equation}
\langle \tilde C_{f_s} \rangle_-=\left[\frac{1-x^2_{f_s}}{1+x^2_{f_s}}
\right]+{\cal O}(r_D^2),
\end{equation}
allowing a straightforward determination of $x_{f_s}$ (see 
(\ref{x-det-LO})). Finally, we may use
\begin{eqnarray}
\eta_{f_s}\langle \tilde S_{f_s} \rangle_-&=&
\left[\frac{2\,x_{f_s}\sin\delta_{f_s}\sin\phi_q}{1+x_{f_s}^2}\right]
\mbox{sgn}(\sin\gamma)-\tilde\Gamma_{f_s}\,\eta_{f_s}\langle \tilde 
S_{f_s}\rangle_+\nonumber\\
&&-2\,r_D\sin\delta_D\cos\phi_q+{\cal O}(r_D^2)\label{Sm-special}
\end{eqnarray}
to determine $\sin\delta_{f_s}$. Employing (\ref{dyn-assumpt}), we obtain 
an unambiguous solution for $\delta_{f_s}$.

In the unlikely case of $\delta_{f_s}=90^\circ \lor 270^\circ$, which 
is the second possibility yielding $\Gamma_{+-}^{f_s}=0$, 
$\tilde\Gamma_{f_s}$ would vanish as well, as can be seen in
(\ref{Gam-NE-1}). Moreover, because of
\begin{equation}
\eta_{f_s}\langle \tilde S_{f_s} \rangle_+=2\,r_D\cos\delta_D
\left[\sin\phi_q+(1-\langle \tilde C_{f_s}\rangle_-)
\cos(\phi_q+\gamma)\sin\gamma\right]+{\cal O}(r_D^2),
\end{equation}
the observable $\langle \tilde S_{f_s} \rangle_+$ would essentially be 
due to ${\cal O}(r_D)$ terms and hence take a small value. On the other 
hand, we have
\begin{eqnarray}
\lefteqn{\eta_{f_s}\langle \tilde S_{f_s} \rangle_-=-
\left[\frac{2\,x_{f_s}\cos(\phi_q+\gamma)}{1+x_{f_s}^2}\right]
\mbox{sgn}(\sin\delta_{f_s})}\nonumber\\
&&-2\,r_D\sin\delta_D\left[\cos\phi_q-(1-\langle \tilde C_{f_s} \rangle_-)
\cos(\phi_q+\gamma)\cos\gamma\right]+{\cal O}(r_D^2)
\end{eqnarray}
\begin{equation}
\langle \tilde C_{f_s} \rangle_-=\left[\frac{1-x^2_{f_s}}{1+x^2_{f_s}}
\right]-2\,r_D\sin\delta_D\left[\frac{2\,x_{f_s}\cos\gamma}{1+x_{f_s}^2}
\right]\mbox{sgn}(\sin\delta_{f_s})+{\cal O}(r_D^2),
\end{equation}
so that these two observables may both take large values. Interestingly,
$r_D$ enters here only in combinations with $\sin\delta_D$, which is
probably small. Consequently, the $r_D$ corrections are expected to play
a very minor r\^ole now. Neglecting them, $x_{f_s}$ and $\phi_q+\gamma$ 
can be straightforwardly determined, even though the latter quantity 
suffers from a fourfold discrete ambiguity. Note that the value of 
$x_{f_s}$ could differ dramatically from $R_b\approx0.4$ in the case of 
$\delta_{f_s}=90^\circ \lor 270^\circ$. Since $\tilde\Gamma_{f_s}$
vanishes for these strong phases, this quantity would no longer allow us
to determine $r_D\sin\delta_D$ and $r_D\cos\delta_D$. However, we may
instead use $\langle \tilde C_{f_s} \rangle_+$, which is given by
\begin{equation}
\langle \tilde C_{f_s} \rangle_+=-2\,r_D\cos\delta_D
\left[\frac{2\,x_{f_s}\sin\gamma}{1+x_{f_s}^2}\right]
\mbox{sgn}(\sin\delta_{f_s})+{\cal O}(r_D^2),
\end{equation}
and implies
\begin{equation}
r_D\cos\delta_D=\left[\frac{\cos(\phi_q+\gamma)
\langle \tilde C_{f_s} \rangle_+}{2\,\sin\gamma\,
\eta_{f_s}\langle \tilde S_{f_s} \rangle_-}\right]
+{\cal O}(r_D^2)
\end{equation}
\begin{equation}
r_D\sin\delta_D=\pm\sqrt{r_D^2-\left[\frac{\cos(\phi_q+\gamma)
\langle \tilde C_{f_s} \rangle_+}{2\,\sin\gamma\,
\eta_{f_s}\langle \tilde S_{f_s} \rangle_-}\right]^2}+{\cal O}(r_D^2).
\end{equation}
Let us emphasize again that the discussion of the
$\delta_{f_s}=90^\circ \lor 270^\circ$ case given above appears to be
rather academic. On the other hand, $\gamma$ may well take a value of
$90^\circ$ or $270^\circ$. It is therefore important that our strategy works
also in this interesting special case.

\boldmath
\subsection{Numerical Examples}\label{sec:examp}
\unboldmath
It is very instructive to illustrate the strategies discussed above by
a few examples. To this end, we use the following hadronic
input parameters:
\begin{equation}
x_{f_s}=0.4, \quad \delta_{f_s}=200^\circ, \quad r_D=0.06, \quad
\delta_D=210^\circ,
\end{equation}
and consider different choices for $\gamma$ and $\phi_q$. Concerning
the strong phases, we assume that $\delta_{f_s}$ and 
$\delta_D$ differ sizeably from their ``na\"\i ve'' values
of $180^\circ$, as we would also like to 
see how their extraction works. In Tables~\ref{tab:illu1} and 
\ref{tab:illu2}, we collect the resulting values 
for the relevant observables.  Here $(\phi_q,\gamma)=(47^\circ,60^\circ)$
and $(133^\circ,120^\circ)$ refer to the $B_d$ case, i.e.\ to the
$B_d\to DK_{\rm S(L)}$ modes; the first parameter set corresponds
to the Standard Model, whereas the second one, which is suggested by a 
recent analysis of the decay $B_d\to\pi^+\pi^-$ \cite{FlMa2}, requires 
new-physics contributions to $B^0_d$--$\overline{B^0_d}$ mixing.
On the other hand, $(\phi_q,\gamma)=(0^\circ,60^\circ)$ corresponds to the 
Standard-Model description of $B_s$-meson decays of the kind 
$B_s\to D\eta^{(')}, D\phi, ...$, which are interesting modes for
hadron colliders.

\begin{table}
\begin{center}
\begin{tabular}{|c||c|c|c|c|}
\hline
$(\phi_q,\gamma)$ & $\eta_{f_s} S_{f_s}$ & $\eta_{f_s}\overline{S}_{f_s}$ 
& $C_{f_s}$ & $\overline{C}_{f_s}$ \\
\hline
\hline
$(47^\circ,60^\circ)$ & $-0.621$ & $-0.714$ & $-0.687$ & $+0.691$ \\
$(133^\circ,120^\circ)$ & $+0.663$ & $+0.471$ & $-0.747$ & $+0.756$ \\
$(0^\circ,60^\circ)$ & $-0.699$ & $-0.401$ & $-0.687$ & $+0.691$ \\
\hline
\end{tabular}
\caption{The observables provided by the $\sin(\Delta M_qt)$ and
$\cos(\Delta M_qt)$ terms of the time-dependent 
$B_q\to D[\to f_{\rm NE}] f_s$ rates in the case of $x_{f_s}=0.4$,
$\delta_{f_s}=200^\circ$, $r_D=0.06$, $\delta_D=210^\circ$ for 
various values of $(\phi_q,\gamma)$.}\label{tab:illu1}
\end{center}
\end{table}

\begin{table}
\begin{center}
\begin{tabular}{|c||c|c|c|c|c|c|}
\hline
$(\phi_q,\gamma)$ & $\Gamma_{+-}^{f_s}$ & $\tilde\Gamma_{f_s}$ &
$\eta_{f_s} \langle \tilde S_{f_s}\rangle_+$ & 
$\eta_{f_s} \langle \tilde S_{f_s}\rangle_-$ &
$\langle \tilde C_{f_s}\rangle_+$ & $\langle \tilde C_{f_s}\rangle_-$\\
\hline
\hline
$(47^\circ,60^\circ)$ & $-0.324$ & $-0.032$ & $-0.667$ & $-0.046$ &
$+0.002$ & $+0.689$\\
$(133^\circ,120^\circ)$ & $+0.324$ & $-0.035$ & $+0.567$ & $-0.096$ &
$+0.004$ & $+0.751$\\
$(0^\circ,60^\circ)$ & $-0.324$ & $-0.032$ & $-0.550$ & $+0.149$ &
$+0.002$ & $+0.689$\\
\hline
\end{tabular}
\caption{Untagged observables and relevant observable combinations
for the extraction of $\gamma$ and hadronic quantities corresponding
to the parameter sets considered in Table~\ref{tab:illu1}.}\label{tab:illu2}
\end{center}
\end{table}

\begin{table}
\begin{center}
\begin{tabular}{|c||c|c|c|c|}
\hline
$(\phi_q,\gamma)$ & $\gamma$ & $\delta_{f_s}$ & $x_{f_s}$ & $\delta_D$ \\
\hline
\hline
$(47^\circ,60^\circ)$ & $59.8^\circ$ ($62.8^\circ$) &
$199.0^\circ$ ($190.9^\circ$) & $0.404$ ($0.429$) & $205.4^\circ$ \\
$(133^\circ,120^\circ)$ & $120.3^\circ$ ($123.8^\circ$) & $198.6^\circ$ 
($215.8^\circ$) & $0.403$ ($0.377$) & $216.7^\circ$ \\
$(0^\circ,60^\circ)$ & $59.9^\circ$ ($59.5^\circ$) & $199.8^\circ$ 
($204.7^\circ$) & $0.403$ ($0.429$) & $209.5^\circ$ \\
\hline
\end{tabular}
\caption{Extracted values of $\gamma$, $\delta_{f_s}$, $x_{f_s}$
and $\delta_D$ corresponding to the observables listed in 
Table~\ref{tab:illu2}. The numbers in brackets are obtained from the 
simple expressions (\ref{tan-gam-LO}), (\ref{tan-del-det-LO}) and 
(\ref{x-det-LO}), whereas the others take into account the $r_D$ effects, 
following the approach discussed in Subsection~\ref{subsec:main-strat}. 
Note the input parameters $\delta_{f_s}=200^\circ$, $x_{f_s}=0.4$,
$\delta_D=210^\circ$.}\label{tab:illu3}
\end{center}
\end{table}

We observe that all observables in Table~\ref{tab:illu1} take large values, 
which is very promising from an experimental point of view. 
Moreover, $\Gamma_{+-}^{f_s}$ and the observable combinations 
$\langle \tilde S_{f_s}\rangle_+$ and $\langle \tilde C_{f_s}\rangle_-$,
which play the key r\^ole for the determination of $\gamma$ and
$x_{f_s}$, respectively, are also favourably large. On the other hand,
the observables $\langle \tilde S_{f_s}\rangle_-$, which allow the extraction
of $\delta_{f_s}$, are suppressed by the small deviation of this phase from 
$180^\circ$. Consequently, in order to be able to probe 
$\delta_{f_s}\sim200^\circ$, the experimental resolution has to reach the 
$0.1$ level, where we then must care about the $r_D$ corrections. In our 
examples, the observables $\tilde\Gamma_{f_s}$, allowing us to probe 
$\delta_D$, take values at the $0.03$ level. For completeness, we have also 
included the values for the observables $\langle \tilde C_{f_s}\rangle_+$, 
which are negligibly small. Fortunately, they are not required for the 
strategy discussed above. 

\begin{figure}
\centerline{\rotate[r]{
\epsfysize=10.9truecm
{\epsffile{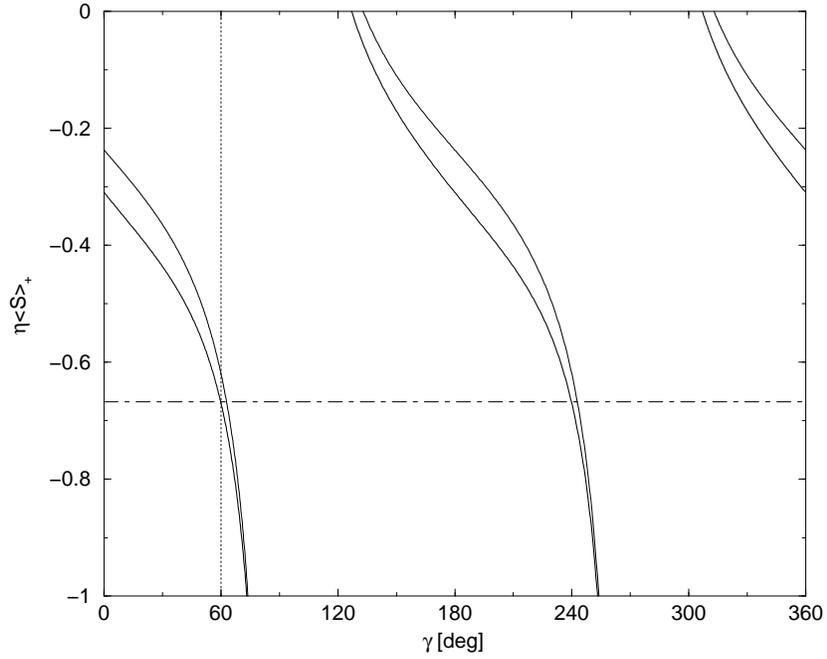}}}}
\caption{The dependence of $\eta_{f_s}\langle \tilde S_{f_s} \rangle_+$ 
on $\gamma$ for the input parameters $\phi_d=47^\circ$ and
$\gamma=60^\circ$, as discussed in the text.}\label{fig:Splus-Bd1}
\end{figure}

The first interesting information on $\gamma$ can already be obtained from 
$\Gamma_{+-}^{f_s}$. Using (\ref{gam-bound}), we obtain 
$-71^\circ\leq\gamma\leq71^\circ$ and $109^\circ\leq\gamma\leq251^\circ$ 
for $\Gamma_{+-}^{f_s}=-0.324$ and $+0.324$, respectively, where we have 
also taken into account (\ref{sgn-cos-gam}). The measurement of 
$\langle \tilde S_{f_s} \rangle_+$ then allows us to {\it determine}
the angle $\gamma$. In Fig.~\ref{fig:Splus-Bd1}, we show the situation
for $(\phi_d,\gamma)=(47^\circ,60^\circ)$. Here the horizontal 
dot-dashed line represents the ``measured'' value of 
$\eta_{f_s}\langle \tilde S_{f_s} \rangle_+$, and the thin solid contours
were calculated with the help of (\ref{S-LO}), neglecting the $r_D$
corrections. We observe that the intersection with the horizontal line 
yields a twofold solution for $\gamma$, which is given by 
$\gamma=63^\circ\lor 243^\circ$. These values can also be obtained directly 
in a simple manner with the help of (\ref{tan-gam-LO}). Using 
(\ref{sgn-cos-gam}), the negative value of $\Gamma_{+-}^{f_s}$ implies 
$\mbox{sgn}(\cos\gamma)=+1$. As $\cos(63^\circ)=+0.45$ and 
$\cos(243^\circ)=-0.45$, we may hence exclude the $\gamma=243^\circ$ 
solution. Alternatively, we could already restrict the range
for $\gamma$ in Fig.~\ref{fig:Splus-Bd1} to $-71^\circ\leq\gamma\leq71^\circ$, 
which corresponds to the $\Gamma_{+-}^{f_s}$ bound (\ref{gam-bound}) discussed
above. However, it is more instructive for our purpose to consider the 
whole $\gamma$ range in this figure. It is interesting to note that
$\gamma=63^\circ$ differs by only $3^\circ$ from our input value of 
$60^\circ$. In order to take into account the $r_D$ corrections, the contours 
in the $\gamma$--$\eta_{f_s}\langle \tilde S_{f_s} \rangle_+$ plane are a
key element, as we have seen above. Using (\ref{S-NLO}), we obtain the 
thick solid contours shown in Fig.~\ref{fig:Splus-Bd1}. Taking again into 
account the $\mbox{sgn}(\cos\gamma)=+1$ constraint, their intersection with 
the horizontal dot-dashed line gives the single solution $\gamma=59.8^\circ$,
which is in perfect agreement with our input value. In Table~\ref{tab:illu3},
we collect also the results for the extracted hadronic parameters. It
is interesting to note that the value of $\delta_D$ determined from
(\ref{rD-sinD}) is in good agreement with our input parameter. Moreover,
the value of $x_{f_s}$ extracted from (\ref{x-det-LO}) agrees also well
with the ``true'' one of 0.4, whereas the value of $\delta_{f_s}$
following from (\ref{tan-del-det-LO}) suffers from large $r_D$
corrections. However, taking them into account through (\ref{SmNLO}), 
we obtain a result that differs by only $1.0^\circ$ from our input value 
of $\delta_{f_s}=200^\circ$. The interesting feature that already the simple 
expressions (\ref{tan-gam-LO}) and (\ref{x-det-LO}) give values of $\gamma$ 
and $x_{f_s}$ that are very close to their ``true'' values is due to the 
fact that the relevant observables, 
$\eta_{f_s}\langle \tilde S_{f_s} \rangle_+=-0.667$ and 
$\langle \tilde C_{f_s} \rangle_-=+0.689$, are much larger than $r_D=0.06$.
On the other hand, this is not the case for 
$\eta_{f_s}\langle \tilde S_{f_s} \rangle_-=-0.046$, which plays the 
key r\^ole for the extraction of $\delta_{f_s}$.

\begin{figure}
\centerline{\rotate[r]{
\epsfysize=10.9truecm
{\epsffile{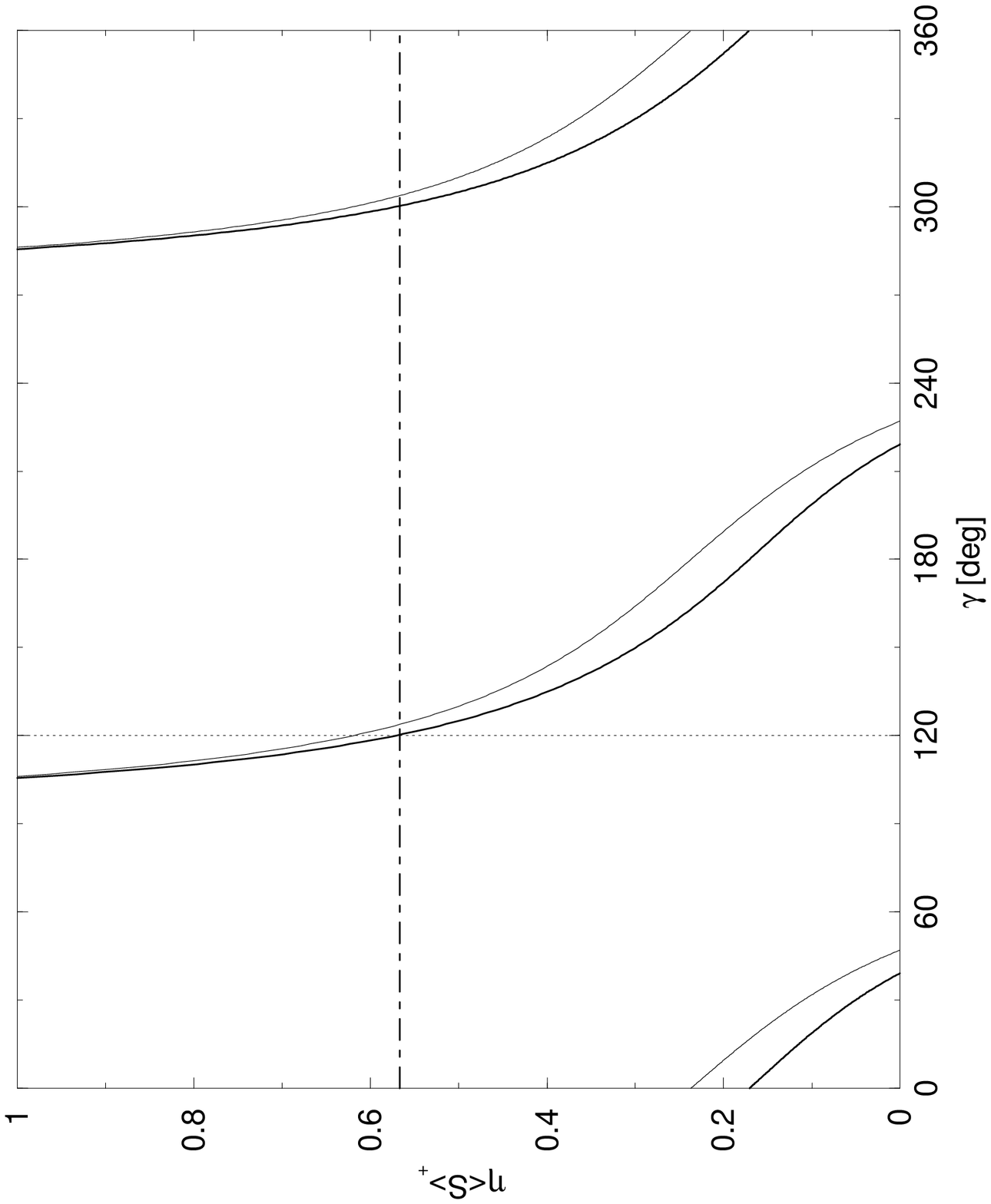}}}}
\caption{The dependence of $\eta_{f_s}\langle \tilde S_{f_s} \rangle_+$ 
on $\gamma$ for the input parameters $\phi_d=133^\circ$ and 
$\gamma=120^\circ$, as discussed in the text.}\label{fig:Splus-Bd2}
\end{figure}

\begin{figure}
\centerline{\rotate[r]{
\epsfysize=10.9truecm
{\epsffile{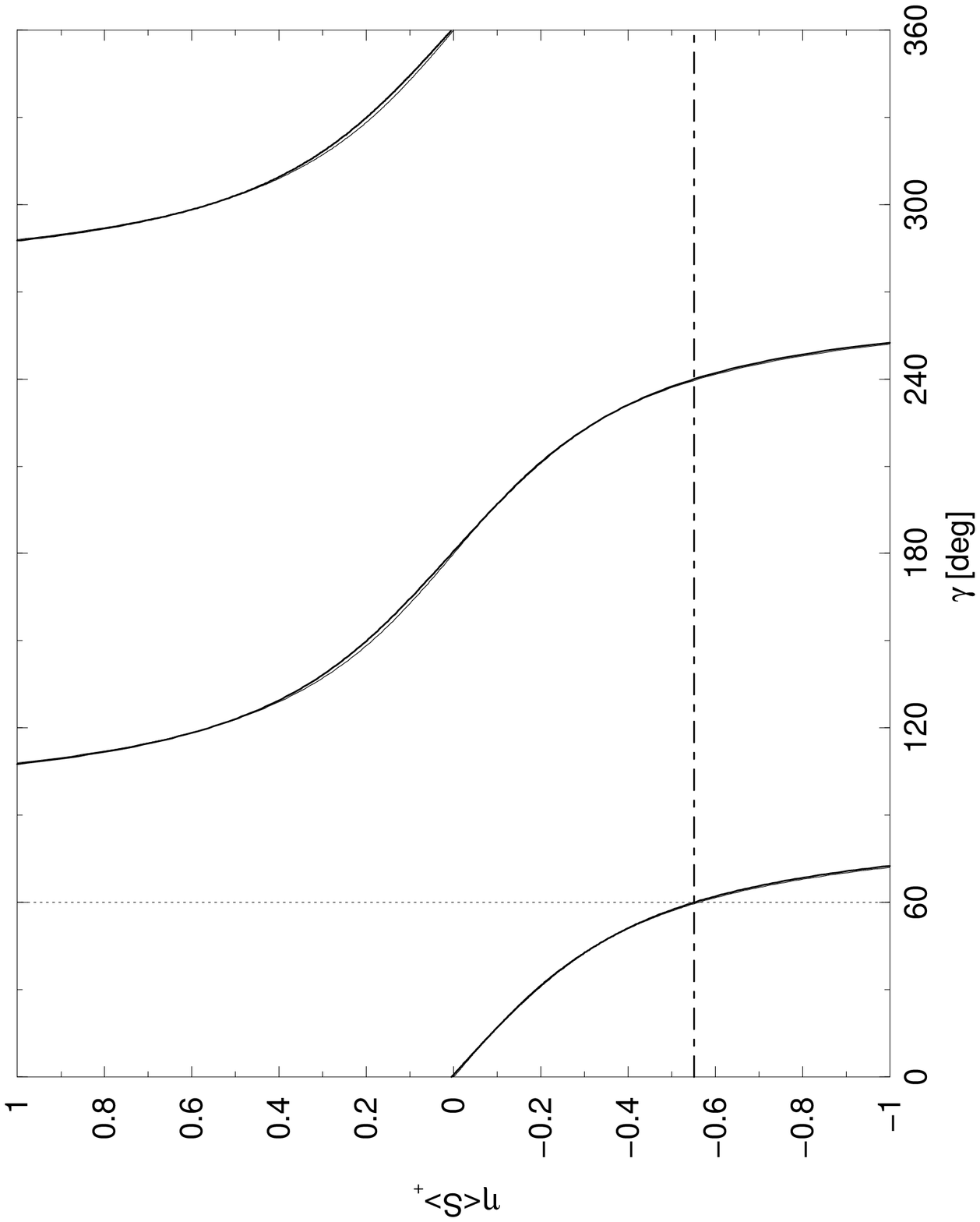}}}}
\caption{The dependence of $\eta_{f_s}\langle \tilde S_{f_s} \rangle_+$
on $\gamma$ for the input parameters $\phi_s=0^\circ$ and $\gamma=60^\circ$, 
as discussed in the text.}\label{fig:Splus-Bs}
\end{figure}

In Figs.\ \ref{fig:Splus-Bd2} and \ref{fig:Splus-Bs}, we repeat this 
exercise for $(\phi_q,\gamma)=(133^\circ,120^\circ)$ and $(0^\circ,60^\circ)$,
respectively. These figures are complemented by the numerical values given 
in Table~\ref{tab:illu3}. The $B_s$-meson case shown in 
Fig.~\ref{fig:Splus-Bs} is particularly interesting, as the $r_D$ effects
would there be essentially negligible, and would also affect, to a smaller 
extent, the extraction of $\delta_{f_s}$. A remarkable feature of the
contours shown in Figs.~\ref{fig:Splus-Bd1}--\ref{fig:Splus-Bs} is that 
already measurements of the $\langle \tilde S_{f_s} \rangle_+$ observables
with rather large uncertainties would allow us to obtain impressive
determinations of $\gamma$.

\boldmath
\subsection{$B_s\to DK_{\rm S(L)}$ and 
$B_d\to D\pi^0, D\rho^0, ...$}\label{sec:r-d-case}
\unboldmath
Let us now finally turn to the $r=d$ case, i.e.\ to decays of the kind
$B_s\to DK_{\rm S(L)}$ and $B_d\to D\pi^0, D\rho^0, ...$, where
the interference effects between the $\overline{B^0_q}\to
\overline{D^0}f_d$ and $\overline{B^0_q}\to D^0f_d$ decay paths are
doubly Cabibbo-suppressed, as reflected by
\begin{equation}\label{x-f-d}
x_{f_d}=-\left(\frac{\lambda^2R_b}{1-\lambda^2}\right)a_{f_d}\approx
-0.02\times a_{f_d}.
\end{equation}
As was pointed out in \cite{RF-BD-CP}, the $B_q\to D_\pm f_d$ case 
provides interesting determinations of 
$\sin\phi_q$, which are -- in comparison with the conventional 
$B_d\to J/\psi K_{\rm S(L)}$ and $B_s\to J/\psi\phi$ methods --
theoretically cleaner by one order of magnitude (see also
Subsection~\ref{subsec:appl-CP}). Let us here focus again on the case 
where the neutral $D$-mesons are observed through decays into CP
non-eigenstates, having in particular $f_{\rm NE}=\pi^+K^-$ in mind. 
Because of (\ref{x-f-d}) and $r_D\approx0.05$,
we keep only the leading-order terms in $x_{f_d}$ and $r_D$, yielding
\begin{eqnarray}
\eta_{f_d}\langle \tilde S_{f_d} \rangle_+&=&\Gamma_{+-}^{f_d}
\left[\sin\phi_q+\tan\gamma\cos\phi_q\right]+2\,r_D\cos\delta_D\sin\phi_q
+...\label{Spd}\\
\eta_{f_d}\langle \tilde S_{f_d} \rangle_-&=-&\Gamma_{+-}^{f_d}
\left[\cos\phi_q-\tan\gamma\sin\phi_q\right]\tan\delta_{f_d}-
2\,r_D\sin\delta_D\cos\phi_q +...,\label{Smd}
\end{eqnarray}
where the dots are an abbreviation for the neglected ${\cal O}(x_{f_d}^2)$, 
${\cal O}(x_{f_d}r_D)$ and ${\cal O}(r_D^2)$ terms. Note that these
observables are expected to take small values, which are not promising from 
an experimental point of view. Moreover, we have
\begin{equation}
\tilde\Gamma_{f_d}=0+...,
\end{equation}
so that this quantity does no longer allow us to fix $r_D\sin\delta_D$ and
$r_D\cos\delta_D$. However, we may instead consider both a $B_s$ and a 
$B_d$ mode, for example $B_s\to D[\to f_{\rm NE}]K_{\rm S(L)}$ and 
$B_d\to D[\to f_{\rm NE}]\rho^0$. Applying (\ref{Spd}) to these channels 
and assuming that $\phi_s$ and $\phi_d$ are known unambiguously, their
observables $\langle \tilde S_{f_d} \rangle_+$ allow us to determine 
straightforwardly $\tan\gamma$ and $r_D\cos\delta_D$. Furthermore, we may 
then apply (\ref{Smd}) to extract the corresponding strong phases 
$\delta_{f_d}$, and may use
\begin{equation}
\Gamma_{+-}^{f_d}=2\,x_{f_d}\cos\delta_{f_d}\cos\gamma +...
\end{equation}
to determine the hadronic parameters $x_{f_d}$. Note that we have
\begin{equation}
\langle \tilde C_{f_d} \rangle_+=0+..., \quad 
\langle \tilde C_{f_d} \rangle_-=1+...,
\end{equation}
so that these observables do not provide useful information in this
case. If $\phi_s$ should take its negligibly small 
Standard-Model value, implying $\sin\phi_s=0$ and $\cos\phi_s=1$, 
(\ref{Spd}) would simplify even further:
\begin{equation}
\left.\eta_{f_d}\langle \tilde S_{f_d} \rangle_+\right|_{\phi_q=0^\circ}
=\Gamma_{+-}^{f_d}\tan\gamma+...,
\end{equation}
thereby allowing a direct determination of $\tan\gamma$. 

Interestingly, because of (\ref{x-f-d}), the $B_q\to D[\to f_{\rm NE}]f_d$ 
observables may be dominated by the $r_D$ terms. In this case, (\ref{Spd}) 
and (\ref{Smd}) would lose their sensitivity on $\gamma$, but would 
instead allow simple determinations of $r_D\cos\delta_D$ and 
$r_D\sin\delta_D$, respectively, thereby yielding $r_D$ and $\delta_D$ 
by using only $B$-decay measurements. These parameters could then also be 
employed as an input for the $B_q\to D[\to f_{\rm NE}]f_s$ analysis 
discussed above.

\boldmath
\section{Conclusions}\label{sec:concl}
\setcounter{equation}{0}
\unboldmath
In this paper, which complements \cite{RF-BD-CP}, we have performed 
a comprehensive analysis of the $B_q\to D f_r$ modes, where the neutral
$D$ mesons may be observed through decays into CP eigenstates $f_\pm$,
or into CP non-eigenstates $f_{\rm NE}$. After a detailed discussion of
the formalism to calculate the observables provided by the
$B_q\to D[\to f_\pm] f_r$ and $B_q\to D[\to f_{\rm NE}] f_r$ transitions, 
we have focused on the latter case, since the exploration of CP violation with 
$B_q\to D_\pm f_r$ modes was already discussed in detail in \cite{RF-BD-CP}. 
Because of interference effects between Cabibbo-favoured and doubly 
Cabibbo-suppressed $D$-meson decays, which are described by a parameter 
$r_D$ determined experimentally from $D$-decay measurements, the situation 
is more complicated in $B_q\to D[\to f_{\rm NE}] f_r$. However, as we 
have pointed out in this paper, we may nevertheless extract $\gamma$ and 
interesting hadronic $B$- and $D$-decay parameters in an efficient and 
essentially unambiguous manner:

\begin{itemize}
\item For the extraction of $\gamma$, the $r=s$ modes, i.e.\ the
$B_d\to DK_{\rm S(L)}$, $B_s\to D\eta^{(')}, D\phi, ...$ channels, 
play the key r\^ole, as certain interference effects between the 
$\overline{B^0_q}\to\overline{D^0}f_s$ and $\overline{B^0_q}\to D^0f_s$ 
decay paths are employed to this end. Whereas these interference effects 
are governed by $R_b\approx0.4$, i.e.\ are favourably large, their 
$r=d$ counterparts are doubly Cabibbo-suppressed. In order to determine 
$\gamma$, we use the $B^0_q$--$\overline{B^0_q}$ mixing phase $\phi_q$ 
as an input, which can be determined through well-known strategies.

\item An interesting bound on $\gamma$ can already be obtained from 
the observable $\Gamma_{+-}^{f_s}$, as 
$|\cos\gamma|\geq|\Gamma_{+-}^{f_s}|$. Complementing $\Gamma_{+-}^{f_s}$ 
through a measurement of the mixing-induced observable 
$\langle\tilde S_{f_s}\rangle_+$, $\gamma$ can be determined in 
an elegant manner with the help of a simple relation.
Measuring, in addition, $\langle\tilde C_{f_s}\rangle_-$, 
we may also extract the hadronic parameter $x_{f_s}$. The theoretical
accuracy of these efficient determinations of $\gamma$ and $x_{f_s}$
is limited by ${\cal O}(r_D)$ corrections. However, since the relevant
observables $\langle\tilde S_{f_s}\rangle_+$ and
$\langle\tilde C_{f_s}\rangle_-$ are expected to be large, which is also 
very promising from an experimental point of view, these effects should 
play a minor r\^ole, as we have also seen in our numerical studies. On 
the other hand, since the strong phase $\delta_{f_s}$ is expected to be 
close to $180^\circ$, its determination from the correspondingly suppressed 
observable $\langle\tilde S_{f_s}\rangle_-$ would require the inclusion of 
the $r_D$ corrections.

\item Following the strategy proposed in this paper, the $r_D$ effects can 
be taken into account straightforwardly with the help of the observables 
$\tilde\Gamma_{f_s}$, $\langle\tilde S_{f_s}\rangle_-$ and 
$\langle\tilde C_{f_s}\rangle_-$. To this end, it is very convenient to
consider contours in the $\gamma$--$\langle\tilde S_{f_s}\rangle_+$ plane.
In the case of our numerical examples, these contours would allow 
impressive determinations of $\gamma$, even if the experimental value for
$\langle\tilde S_{f_s}\rangle_+$ should suffer from a sizeable uncertainty, 
and the $r_D$ corrections would have a small impact. The observables 
$\langle\tilde C_{f_s}\rangle_-$, $\langle\tilde S_{f_s}\rangle_-$ and
$\tilde\Gamma_{f_s}$ provide interesting extractions of $x_{f_s}$, 
$\delta_{f_s}$ and $\delta_D$. The numerical studies suggest, furthermore, 
that the $B_s$-meson decays $B_s\to D\eta^{(')}, D\phi, ...$\ may be 
particularly promising, thereby representing a nice new playground for 
hadronic $B$ experiments of the LHC era. Moreover, there are various
interesting ways to combine the information provided by $B_s$- and 
$B_d$-meson decays.

\item If we assume that $\phi_q$ will be known unambiguously by the time 
these measurements can be performed in practice, our strategy gives a 
twofold solution $\gamma=\gamma_1\lor\gamma_2$, with  
$\gamma_1\in[0^\circ,180^\circ]$ and $\gamma_2\in[180^\circ,360^\circ]$. 
Using the plausible assumption that $\cos\delta_{f_s}<0$, the sign of
$\Gamma_{+-}^{f_s}$ allows us to distinguish between these solutions,
thereby fixing $\gamma$ {\it unambiguously}. In particular, we may check
whether this angle is actually smaller than $180^\circ$, as implied by
the Standard-Model interpretation of $\varepsilon_K$. On the other hand,
if we assume -- as is usually done -- that $\gamma\in[0^\circ,180^\circ]$,
in accordance with $\varepsilon_K$, a knowledge of $\sin\phi_q$ is sufficient
for our strategy. The assumption of $\cos\delta_{f_s}<0$ then allows us 
to determine the sign of $\cos\phi_q$, thereby fixing $\phi_q$
unambiguously, and to obtain also an unambiguous value of $\gamma$.
Following these lines, we could distinguish between the two solutions
$(\phi_d,\gamma)\sim(47^\circ,60^\circ)$ and $(133^\circ,120^\circ)$, which
are suggested by a recent analysis of CP violation in $B_d\to\pi^+\pi^-$. 
Similar comments apply to the $B_q\to D_\pm f_s$ analysis performed in 
\cite{RF-BD-CP}.
 
\item Although the $r=d$ modes $B_s\to D_\pm K_{\rm S(L)}$ and 
$B_d\to D_\pm\pi^0, D_\pm\rho^0, ...$\ provide interesting 
determinations of $\sin\phi_s$ and $\sin\phi_d$, respectively, as pointed 
out in \cite{RF-BD-CP}, decays of this kind are not very attractive for the 
determination of $\gamma$. In the case of $B_q\to D[\to f_{\rm NE}] f_d$, 
we have furthermore to deal with the $r_D$ effects. It is nevertheless 
important to make efforts to measure the corresponding observables. In 
particular, they may even be dominated by the ${\cal O}(r_D)$ terms, 
and would then allow us to extract $r_D$ and $\delta_D$ in a very simple 
manner. These quantities could then be used as an input for the extraction
of $\gamma$ from the $B_q\to D[\to f_{\rm NE}] f_s$ channels.
\end{itemize}
Following these lines, $B_q\to Df_r$ decays provide very interesting 
tools to explore CP violation. Moreover, we may obtain valuable insights 
into the hadron dynamics of the corresponding $B$- and $D$-meson decays. 
Consequently, we strongly encourage detailed experimental studies. 
Since we are dealing with colour-suppressed $B$ decays, exhibiting 
branching ratios at the $10^{-5}$ level in the $r=s$ case, these strategies 
appear to be particularly interesting for the next generation of dedicated 
$B$ experiments, LHCb and BTeV, as well as those at super-$B$ factories.
However, first important steps may already be achieved at the present 
$B$-decay facilities, i.e.\ at BaBar, Belle and run II of the Tevatron.

\end{document}